\documentclass[10pt,journal,compsoc]{IEEEtran}

\ifCLASSOPTIONcompsoc
  \usepackage[nocompress]{cite}
\else
  \usepackage{cite}
\fi

\usepackage{natbib}
\usepackage{amsmath,amsfonts}
\usepackage{algorithm}
\usepackage{algpseudocode}
\usepackage{enumitem}
\usepackage{array}
\usepackage{textcomp}
\usepackage{stfloats}
\usepackage{url}
\usepackage{verbatim}
\usepackage{graphicx}
\usepackage{booktabs}
\usepackage{xcolor}
\usepackage{color}
\usepackage{makecell}
\usepackage{caption}
\usepackage{multirow}
\def\BibTeX{{\rm B\kern-.05em{\sc i\kern-.025em b}\kern-.08em
    T\kern-.1667em\lower.7ex\hbox{E}\kern-.125emX}}
\usepackage{balance}
\usepackage{subfigure}
\usepackage{ragged2e}

\newcommand{\ie}{\textit{i}.\textit{e}.}
\newcommand{\eg}{\textit{e}.\textit{g}.}

\newcommand{\etc}{\textit{etc}}

\newcommand{\revise}[1]{{#1}}

\usepackage{scalerel}
\usepackage{tikz}
\usetikzlibrary{svg.path}
\usepackage[pagebackref=false,breaklinks=true,colorlinks,bookmarks=false]{hyperref}

\definecolor{orcidlogocol}{HTML}{A6CE39}
\tikzset{
  orcidlogo/.pic={
    \fill[orcidlogocol] svg{M256,128c0,70.7-57.3,128-128,128C57.3,256,0,198.7,0,128C0,57.3,57.3,0,128,0C198.7,0,256,57.3,256,128z};
    \fill[white] svg{M86.3,186.2H70.9V79.1h15.4v48.4V186.2z}
                 svg{M108.9,79.1h41.6c39.6,0,57,28.3,57,53.6c0,27.5-21.5,53.6-56.8,53.6h-41.8V79.1z M124.3,172.4h24.5c34.9,0,42.9-26.5,42.9-39.7c0-21.5-13.7-39.7-43.7-39.7h-23.7V172.4z}
                 svg{M88.7,56.8c0,5.5-4.5,10.1-10.1,10.1c-5.6,0-10.1-4.6-10.1-10.1c0-5.6,4.5-10.1,10.1-10.1C84.2,46.7,88.7,51.3,88.7,56.8z};
  }
}

\newcommand\orcidicon[1]{\href{https://orcid.org/#1}{\mbox{\scalerel*{
\begin{tikzpicture}[yscale=-1,transform shape]
\pic{orcidlogo};
\end{tikzpicture}
}{|}}}}

\begin{document}

\title{ENCODE: Breaking the Trade-off Between Performance and Efficiency in Long-term User Behavior Modeling}

\author{
    Wen-Ji Zhou$^*$ \orcidicon{0000-0002-5116-5679},
    Yuhang~Zheng$^*$ \orcidicon{0000-0001-9628-1940},
    Yinfu Feng$^\dagger$ \orcidicon{0000-0001-9136-0965},
    Yunan Ye \orcidicon{0000-0002-0891-5040},
    Rong Xiao \orcidicon{orcid},
    Long Chen \orcidicon{0000-0001-6148-9709},
    Xiaosong Yang \orcidicon{orcid},
    and Jun Xiao \orcidicon{0000-0002-6142-9914}
     \\
    \IEEEcompsocitemizethanks{
        \IEEEcompsocthanksitem $^*$ These authors contributed equally, and $^\dagger$ denotes corresponding author. 
        
        \IEEEcompsocthanksitem Wen-Ji Zhou, Yuhang~Zheng, Yunan Ye, Yinfu Feng, and Rong Xiao are with the Alibaba Group, Hangzhou, China, 310000. Email: \{eric.zwj, zhengyuhang.zyh,yunan.yyn\}@alibaba-inc.com, fyf200502@gmail.com, xiaorong.xr@taobao.com, 

        \IEEEcompsocthanksitem Long Chen is with the Department of Computer Science and Engineering, The Hong Kong University of Science and Technology, Kowloon, Hong Kong SAR, 999077. Email: longchen@ust.hk.

        \IEEEcompsocthanksitem Xiaosong Yang is with the Bournemouth University, Bournemouth, England. Email: xyang@bournemouth.ac.uk.

        \IEEEcompsocthanksitem Jun Xiao is with the Zhejiang University, Zhejiang, China. Email: junx@cs.zju.edu.cn.
    }
    \thanks{Manuscript received 29 March 2024; revised 3 September 2024; accepted 17 October 2024.}
}

\markboth{Journal of IEEE Transactions on Knowledge and Data Engineering,Vol.~,No.~,April 2024}%
{ENCODE: Breaking the Trade-off Between Performance and Efficiency in Long-Term User Behavior Modeling}

\IEEEtitleabstractindextext{%
    \justifying 
    \begin{abstract}
        Long-term user behavior sequences are a goldmine for businesses to explore users’ interests to improve Click-Through Rate (CTR). However, it is very challenging to accurately capture users' long-term interests from their long-term behavior sequences and give quick responses from the online serving systems. To meet such requirements, existing methods “inadvertently” destroy two basic requirements in long-term sequence modeling:
        \textbf{R1}) make full use of the entire sequence to keep the information as much as possible;~\textbf{R2}) extract information from the most relevant behaviors to keep high relevance between learned interests and current target items. The performance of online serving systems is significantly affected by incomplete and inaccurate user interest information obtained by existing methods. To this end, we propose an efficient two-stage long-term sequence modeling approach, named as \textbf{E}fficie\textbf{N}t \textbf{C}lustering based tw\textbf{O}-stage interest mo\textbf{DE}ling (ENCODE), consisting of offline extraction stage and online inference stage. It not only meets the aforementioned two basic requirements but also achieves a desirable balance between online service efficiency and precision. Specifically, in the offline extraction stage, ENCODE clusters the entire behavior sequence and extracts accurate interests. To reduce the overhead of the clustering process, we design a metric learning-based dimension reduction algorithm that preserves the relative pairwise distances of behaviors in the new feature space. While in the online inference stage, ENCODE takes the off-the-shelf user interests to predict the associations with target items. Besides, to further ensure the relevance between user interests and target items, we adopt the same relevance metric throughout the whole pipeline of ENCODE.
        The extensive experiment and comparison with SOTA on both industrial and public datasets have demonstrated the effectiveness and efficiency of our proposed ENCODE.
    \end{abstract}
	
    \begin{IEEEkeywords}
        CTR prediction, user behavior modeling, metric learning, recommendation systems
    \end{IEEEkeywords}
}
\maketitle

\section{Introduction}
\label{sec:intro}
\IEEEPARstart{B}{ecause} of the huge commercial value, Click-Through Rate (CTR) prediction models have received great attention from both academia and industry in recent years. To capture the intrinsic interests of users, user behavior sequences that can comprehensively reflect users' interests have been introduced in CTR modeling and gained encouraging results~\cite{pi2019practice,pi2020search,Qin0WJF020,wu2020sse,de2021transformers4rec}.
Meanwhile, with the rapid development of the Internet and smartphone technologies, users are becoming increasingly engaged with e-commerce websites, leading to an enormous surge of user behavior data. As pointed out in the work~\cite{RenQF0ZBZXYZG19}, 23\% of the users in some e-commerce websites had more than 1,000 clicks over the past 5 months and the accumulated user behavior sequences on mature Internet service platforms have become extremely long since the user's first registration. Therefore, it is feasible to set up long-term user behavior sequences in practice~\cite{zhou2021large,hansen2020contextual,TaoLLCWCLL22,XuDLHL20}.

As user behavior sequences become longer, it becomes important to leverage the abundant information contained within them effectively and efficiently~\cite{zhang2021deep}. Numerous studies have focused on long-term user behavior sequence modeling, but achieving a balance between precision and efficiency in prediction models remains challenging due to strict latency requirements in many online systems.

Roughly speaking, existing long-term modeling strategies can be categorized into two groups: online search-based methods~\cite{pi2020search,Qin0WJF020} and offline interest modeling-based methods~\cite{ren2019lifelong,pi2019practice,cao2022sampling}. \textit{(1) Online search-based methods:} these methods first roughly search for the top-$k$ most relevant items\footnote{To avoid any confusion, the term “item” in this paper refers to the product in e-commerce.} from the user behavior sequence concerning target items, then perform complex advanced modeling algorithms on these retrieved items. To search for relevant items efficiently, people often depend on additional information about the items or use a rough similarity measure between the items and the user's behavior. This helps them establish a
relevance metric for the search. However, the value of $k$, which is the number
of items retrieved, is often limited to less than 100, which impedes the full
utilization of the entire user behavior sequence.
\textit{(2) Offline interest modeling-based methods:} These methods tend to pre-process users' entire long-term sequences into several off-the-shelf embeddings for facilitating online inference. Due to the lack of guidance from target items during the interest modeling stage, it can not ensure high relevance between these offline-generated users' interest representations (\ie, embeddings) and their corresponding target items.

To meet the latency requirements of online service systems, existing methods ``inadvertently" destroy two basic requirements in long-term sequence modeling: \textit{\textbf{R1)} make full use of the entire sequence to keep the information as much as possible.} The entire user behavior sequence should be exploited to extract users' long-term and stable interests. For online search-based methods, it is difficult to meet this requirement, especially for those users with rich historical behaviors. If the number of related historical behaviors to the target items is much greater than $k$~(\eg, it is very common in some categories like the clothing category in most e-commerce platforms), then searching for $k$ behaviors will cause the retrieved behaviors can not fully characterize users' interests leading to affect the performance of CTR models.
\textit{\textbf{R2)} Extract information from the most relevant behaviors to keep high relevance between learned interests and current target items.}
To ensure high relevance related to current target items, the long-term interests should be extracted from the most relevant historical behaviors.
There are two reasons why existing methods have failed to ensure relevance between learned interests and target items: 1) lack of guidance from target items~(\eg, offline interest modeling-based methods); and 2) inconsistent use of relevance metrics in different processing steps, resulting in the information loss about relevant behaviors in some steps which takes the coarse-grained relevance metric.
Take online search-based methods as an example, the coarse-grained relevance metric~(\eg, category) used for searching would lead to fetching out irrelevant historical behaviors and reducing the algorithm performance. 

\begin{figure}
    \centering
    \includegraphics[width=0.95\linewidth]{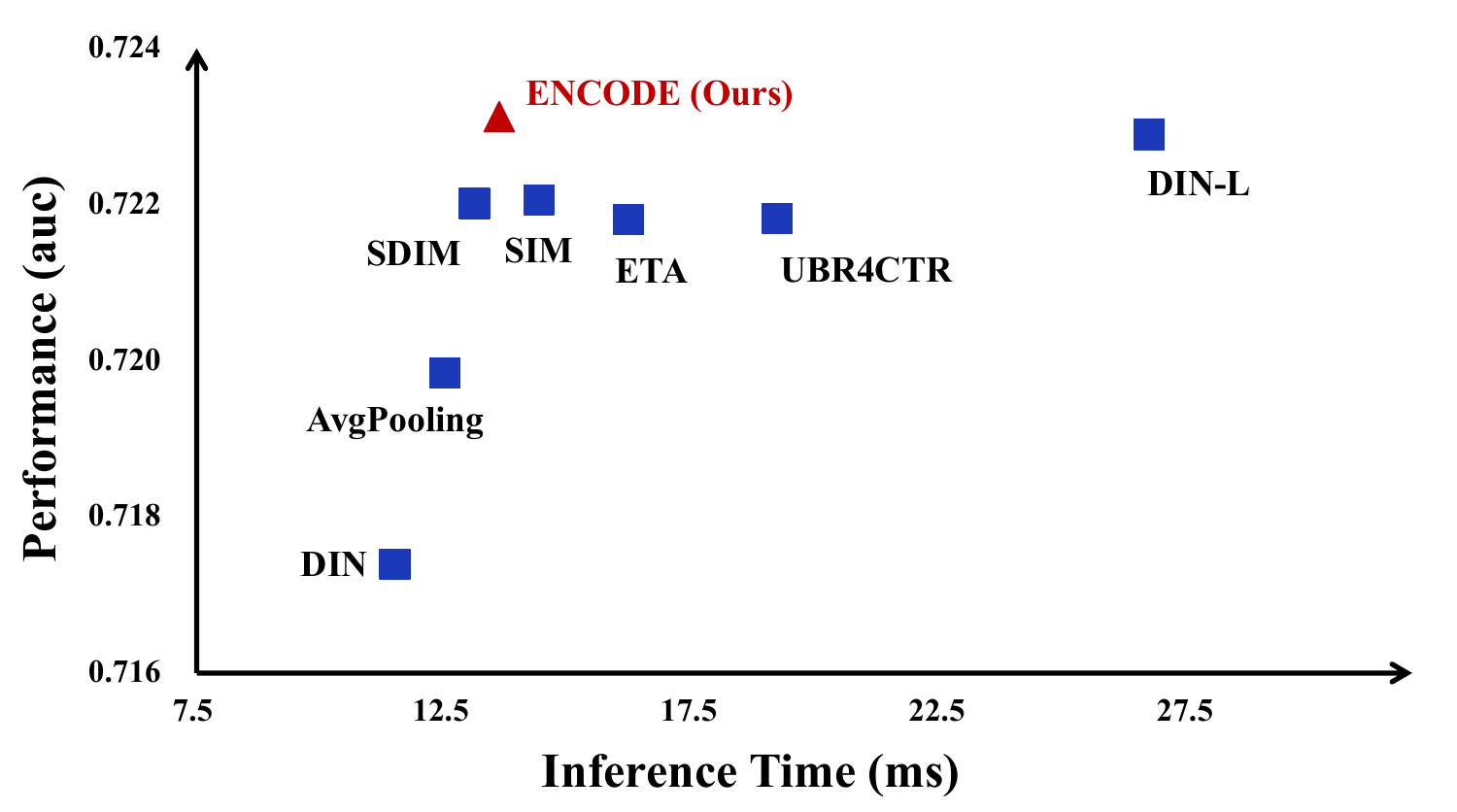}
    \vspace{-1em}
    \caption{Performance and inference time comparison with some SOTA methods for long-term sequence modeling.}

    \vspace{-1em}

  \label{fig:intro}
\end{figure}

In this paper, we propose an efficient two-stage long-term sequence modeling approach for CTR prediction, named as \textbf{E}fficie\textbf{N}t \textbf{C}lustering based tw\textbf{O}-stage interest mo\textbf{DE}ling~(ENCODE). Overall, ENCODE includes an offline extraction stage to fully extract the user's interests and an online inference stage to infer their interests concerning target items. 
During the offline stage, ENCODE first applies clustering algorithms to group the user's entire long-term sequence into multiple clusters. Then within each cluster, a target attention module is utilized to extract the cluster representation, effectively characterizing users' interests. In this way, the original lengthy sequence is condensed into a set of user's sub-interests for subsequent online inference. Meanwhile, we develop a metric learning-based dimension reduction algorithm to reduce clustering overhead and preserve relative pairwise distances of behaviors in the new embedding feature space, enabling offline devices to handle the computational workload.
To well exploit the guidance information of target items for CTR prediction, we design an interest inference module in an online system to obtain the final long behavior sequence representation using a target attention approach.
By combining online and offline attention modules, ENCODE models the entire behavior sequence while ensuring the relevance between extracted long-term interests and target items, satisfying \textbf{R1} and \textbf{R2}.
Furthermore, ENCODE uses the same metric function in both online and offline stages, ensuring high relevance between extracted interests and target items (\ie,\textbf{R2}). In practical deployment, ENCODE periodically extracts users' long-term interests from the complete user sequence to offer a more personalized experience that captures changes in user interests.

ENCODE outperforms most state-of-the-art long-term sequence modeling methods. It achieves superior performance and faster inference speed, as shown in Fig.~\ref{fig:intro}. The main contributions are summarized as below: 

\begin{itemize} [left=1em]
  \item We propose ENCODE, a novel two-stage approach for long-term sequence modeling that meets the above requirements. ENCODE outperforms existing SOTA methods in both performance and effectiveness.

  \item We develop a metric learning-based dimension reduction algorithm for fast clustering, which not only effectively preserves the relative pairwise distances of the behaviors in the new embedding feature space but also significantly reduces the overhead involved in subsequent clustering tasks.

  \item  We validate the effectiveness of ENCODE through offline experiments conducted on both public and industrial datasets. Furthermore, we conducted thorough ablation studies to demonstrate the importance of meeting both \textbf{R1} and \textbf{R2} to achieve improved performance.
  
\end{itemize}

\section{Related Works}

\subsection{Click-Through Rate~(CTR) Prediction}
Click-through rate (CTR) prediction is a critical task in e-commerce scenarios wherein it is widely used in search, recommendation, and advertising. CTR models predict the likelihood that something on a website (such as a product or advertisement) will be clicked by the user, and the outputs can be used as a ranking score for downstream-related tasks. The existing work on CTR prediction related to our research topic can be divided into two technical directions: \textbf{feature interaction} and \textbf{user interest modeling}.

\textbf{Feature Interaction}: The main objective of feature interaction in CTR prediction is to capture the intricate relationships between features, which can enhance the accuracy of prediction models. These methods usually learn and exploit the co-occurrence of feature pairs and labels. Representative works in this field include FM~\cite{Rendle10}, FFM~\cite{PanXRZPSL18}, Wide\&Deep~\cite{Cheng0HSCAACCIA16}, AFM~\cite{XiaoY0ZWC17}, DeepFM~\cite{GuoTYLH17}, PNN~\cite{QuCRZYWW16}, xDeepFM~\cite{LianZZCXS18}, \etc. In addition, some methods, such as DCN~\cite{wang2017dcn}, try to model the relationship between higher-order feature combinations and click labels. These methods can be used in conjunction with user behavior sequences to improve prediction accuracy in real-world applications.

\begin{figure*}[t]
    \centering
    \includegraphics[width=0.97\linewidth]{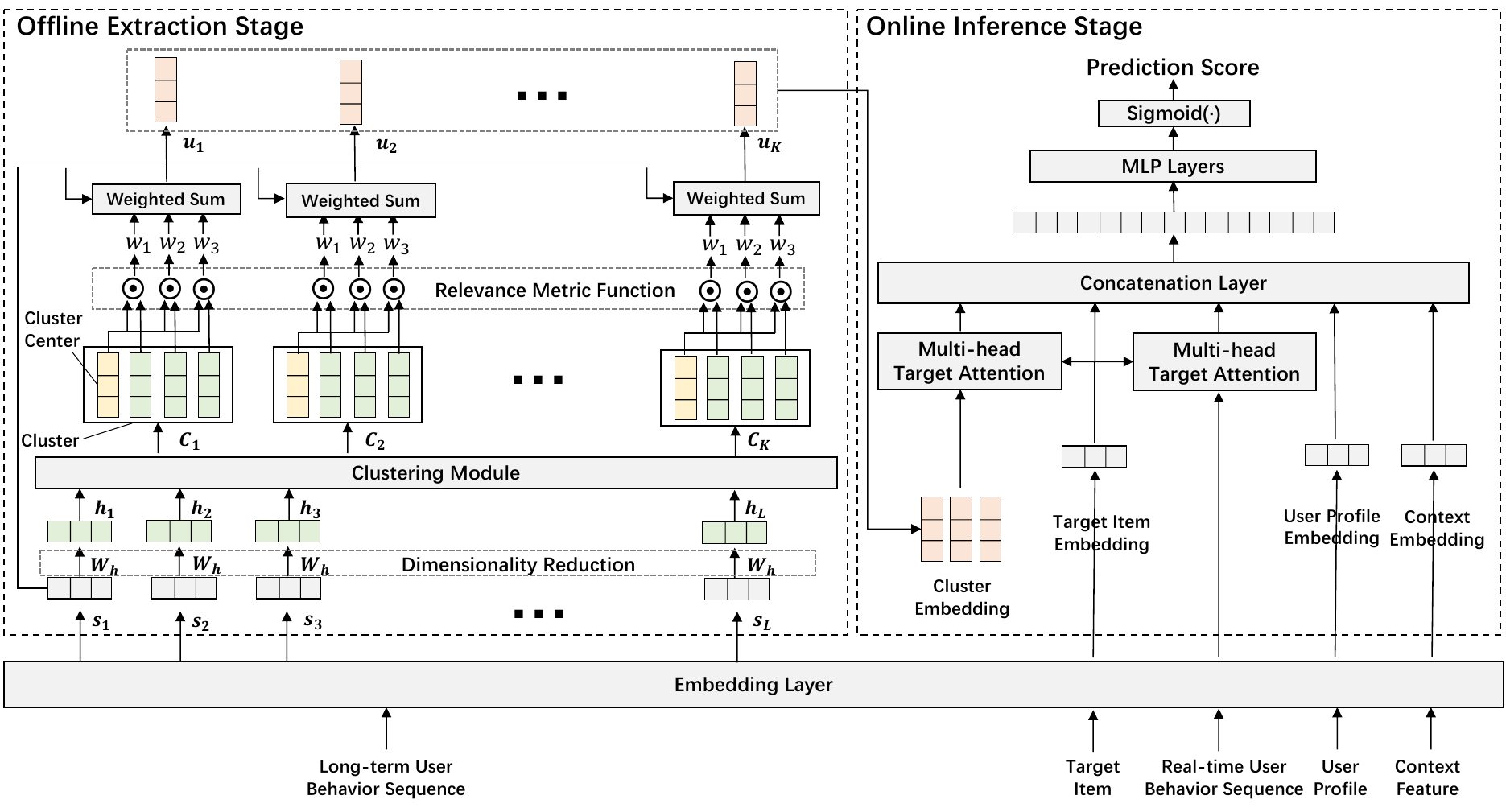}
    \caption{The pipeline of our proposed ENCODE which consists of two stages: \textbf{(1) offline extraction stage:} ENCODE first reduces the dimension of the behavior sequence, then clusters the sequence into several clusters, and finally extracts cluster representations as the user's multi-interests. \textbf{(2) Online inference stage:} ENCODE takes the offline cluster representations and other features as input to predict the probability of clicking on the target item~(\ie, CTR).} 
\label{fig:encode}
\end{figure*}

\textbf{User Interest Modeling}: User interest modeling is an essential technique that learns user interests from their historical behavioral sequences to achieve highly personalized CTR prediction. 
One example of its successful application is YoutubeDNN~\cite{CovingtonAS16}, which feeds the user's video browsing sequence in the model and gains enhanced prediction accuracy. 
\revise{According to the model architecture, these methods can be categorized into RNN-based methods~\cite{HidasiKBT15,ZhouMFPBZZG19} and Attention-based methods~\cite{FengLSWSZY19,SunLWPLOJ19,zhou2018deep}. 
Recently, the mainstream approach in industry applications is to extract multiple interests from user history behaviors to retrieve candidates from diverse fields, such as  MIND~\cite{li2019multi}, MTIN~\cite{jiang2020aspect}, ComiRec~\cite{cen2020controllable}, \textit{etc}.}
\revise{Unfortunately, the classic user interest modeling methods are not directly applicable to long sequence modeling due to their high time complexity.}

\subsection{Long-Term User Behavior Sequence Modeling}
Then, we briefly introduce the above-mentioned two classes of long-term user behavior sequence modeling methods.

\textbf{Online search-based methods}. These methods including UBR4CTR~\cite{Qin0WJF020}, SIM~\cite{pi2020search}, ETA~\cite{chen2022efficient} and TWIN~\cite{chang2023twin} leverage target item information to search for the top-$k$ most relevant items from user behavior sequences. In UBR4CTR and SIM, relevant items are retrieved according to the target item's attributes~(\eg, category). To achieve more accurate retrieval, ETA adopts the LSH algorithm and Hamming distance to retrieve embedding-similar behaviors. For efficiency, the coarse-grained relevance metric is used for searching and the value of $k$ can only be up to 100. However, these methods fail to meet the basic requirements \textbf{R1} and \textbf{R2}. Fortunately, a recent work called TWIN~\cite{chang2023twin} also recognizes the issue of inconsistent use of relevance metrics in different stages and attempts to address it by using the same relevance metric function across all steps. But TWIN still faces a limitation in terms of the length $k$.

\textbf{Offline interest modeling-based methods}. These methods tend to adopt sophisticated modeling algorithms that extract users' multi-interests from long-term sequences in advance for facilitating online inference, primarily including memory-based methods and sampling-based methods. Memory-based methods, such as MIMN~\cite{ren2019lifelong}, update interest extraction modules~(typically based on Recurrent Neural Network~(RNN)) in real-time but may struggle to predict CTR during significant changes in user interest. Meanwhile, due to the inherent inadequacy of RNN, these methods tend to forget the user's earlier interests, violating \textbf{R1}. Moreover, the absence of guidance from the target item during interest modeling also makes it challenging to ensure relevance to the target item, failing to meet \textbf{R2}. To overcome these limitations, SDIM~\cite{cao2022sampling} is proposed as a sampling-based method that uses an offline hash module to generate hash signatures for each historical behavior and approximates attention weights with probability of hash collisions. Although SDIM considers the entire sequence of behavior, a gap exists between the probability of hash collision and attention weights, which results in failing to meet R2.


\section{Methodology}



Following the same conventions of existing works, the CTR prediction task can be formulated as a binary classification problem. Given a training dataset $\mathcal{D} = \{x_i, y_i\}_{i=1}^{N}$ consisting of input features $x_i \in \mathbb{R}^{d_x} $ and the corresponding ground-truth labels $y_i \in \{0,1\}$, the CTR model learns a complex mapper $f: \mathbb{R}^{d_x} \rightarrow \mathbb{R}$ to perform CTR prediction. The predicted CTR $\hat{y_i}$ is calculated as below:
\begin{equation}
    \hat{y_i} = \sigma(f(x_i)),
\end{equation}
where $\sigma(\cdot)$ is the sigmoid function. Usually, the CTR model is trained by minimizing the binary cross-entropy (CE) loss:
\begin{equation}
    \mathcal{L}_{CE} = -\frac{1}{N} \sum_{i=1}^{N} y_i\log(\hat{y_i}) + (1 - y_i) \log(1 - \hat{y_i}).
\end{equation}
For simplicity, we omit subscript $i$ in the following parts.

As shown in Fig.~\ref{fig:encode}, the input feature $x$ usually consists of five groups: user profile feature $x_{user}$, context feature $x_{ctx}$, target item feature $x_t$, real-time user behavior sequence $\mathbf{r} = \{r_i\}_{i=1}^M$ and long-term user behavior sequence $\mathbf{s} = \{s_i\}_{i=1}^L$, where $M$ is the length of real-time user behavior sequences~(\eg, $M < 100$), $L$ is the length of long-term user behavior sequences~(\eg, $L > 500$). \revise{For the embedding layer, we divide the input features into categorical features and continuous features. Specifically, for categorical features, we employ embeddings to transform sparse categorical data into dense representations that are compatible with neural networks, with each unique ID having its own learned embedding. For continuous features, we apply appropriate normalization to ensure that the values are scaled and uniformly distributed within the range of [0, 1).}

\revise{After embedding layer, existing CTR models usually utilize multiple multilayer perceptron (MLP) to process the first three groups of features, \textit{i.e.}, $x_{user}$, $x_{ctx}$, and $x_t$. } For real-time user behavior sequences $\mathbf{r}$, standard target attention~\cite{zhou2018deep} is adopted. While for the last long-term user behavior sequences $\mathbf{s}$, search-based and sampling-based methods are widely used.

In the rest of this section, we first compare the proposed ENCODE with existing long-term user behavior sequence modeling methods~(Subsection~\ref{methods:cmp}). Then we introduce details of ENCODE, including the offline extraction stage~(Subsection~\ref{sec:offline}) and online inference stage~(Subsection~\ref{sec:online}).

\subsection{ENCODE vs. Existing Methods \label{methods:cmp}} 
To further highlight the advantages of our ENCODE, we compare ENCODE with mainstream online search-based and offline interest modeling-based methods. For offline interest modeling-based methods, we just compare sampling-based methods since memory-based methods are often affected by recent behaviors resulting in poor performance.

\textbf{Online search-based Methods.} Given a long-term user behavior sequence $\mathbf{s}$=$\{s_i\}_{i=1}^L$, search-based methods roughly retrieve top-$k$ most relevant items $\mathbf{s}^r$=$\{s^r_i\}_{i=1}^k$ and then apply standard target attention to them. To meet the latency requirements of online systems, $k$ is almost restricted to less than 100. In other words, these methods have to discard another $L-k$ user behaviors. Although discarded behaviors are less relevant, such a design still causes a certain degree of information loss, especially for users with rich historical behaviors, so does not satisfy \textbf{R1}. Moreover, due to the coarse relevance metric used in the search process~\cite{chang2023twin}, these search-based methods can not guarantee the relevance between search results and target items~(\ie, does not satisfy \textbf{R2}). For example, SIM(hard)~\cite{pi2020search} simply searches same-category items as results but ignores the fact that relevant items may belong to different categories. 

\textbf{Sampling-based Methods.} Given long-term user behavior sequence $\mathbf{s} = \{s_i\}_{i=1}^L$, sampling-based methods try to figure out a set of sampling probability $\mathbf{p} = \{p_i\}_{i=1}^L$ according to the approximate relevance between user interests and target items, then obtain user's interest representations by $\mathbf{I} = \sum_{i=1}^L p_i \cdot s_i$. For example, SDIM~\cite{cao2022sampling} calculates the locality-sensitive hash (LSH)~\cite{alexandr2015practical} of user behavior item embeddings offline in advance, and takes LSH collision probability of target item as the sampling probability online, aiming to approximate the weights calculated by target attention. 
It's inspiring that SDIM utilizes the entire behavior sequence without discarding any behaviors. However, there still exists a gap between LSH collision probability and attention weights, which is primarily due to the failure to meet the requirement of \textbf{R2}. 

\textbf{Proposed ENCODE.} To meet the aforementioned requirements, ENCODE adopts a two-stage approach to model user's long-term historical behaviors: 1) offline extraction stage: given a long-term user behavior sequence $\mathbf{s} = \{s_i\}_{i=1}^L$, ENCODE clusters the behavior sequence into $K$ clusters, and get $K$ cluster centers $\mathbf{c} = \{c_i\}_{i=1}^K$. Subsequently, a target attention module is employed within each cluster to extract more accurate sub-interests $\mathbf{U} = \{u_i\}_{i=1}^K$ by taking the cluster center as the query. 2) online inference stage: to enforce the user's interest representation $\mathbf{I}$ relative to the target item, another target attention module is used by taking target item $x_t$ as query and taking offline multi-interests $\mathbf{U}$ as key and value.
Moreover, ENCODE ensures consistency by employing the same relevance metric throughout both the online and offline stages, which fulfills \textbf{R2}.
By incorporating both online and offline attention modules, ENCODE effectively models the entire behavior sequence while maintaining the relevance between the extracted interests and the target items, satisfying both \textbf{R1} and \textbf{R2}. Next, we detailed introduce these two stages.


\subsection{Offline Multi-Interests Extraction\label{sec:offline}}

Usually, a user's interests are composed of multiple aspects of sub-interests while each sub-interest is reflected in a set of similar historical behaviors. Thus, a straightforward idea to fully model users' interests is clustering behavior sequences into several clusters and then obtaining sub-interest representations from these clusters. 

To this end, given a long-term user behavior sequence $\mathbf{s}$=$\{s_i\}_{i=1}^L$, ENCODE obtains user's multi-interest representations by three main steps: (1) Reduce the dimensionality of $\mathbf{s}$ to a lower dimension to reduce the complexity of subsequent calculations, denoted as $\mathbf{H}$=$\{h_i\}_{i=1}^L$, \ie, line 3-8 in Algo.~\ref{alg:encode}. (2) Cluster $\mathbf{H}$ into $K$ clusters, \ie, line 11 in Algo.~\ref{alg:encode}. (3) Extract clusters' embeddings $\mathbf{U}$=$\{u_i\}_{i=1}^K$ as user's multi-interest representations, \ie, line 14-16 in Algo.~\ref{alg:encode}.

In the following, we introduce the details step by step.

\begin{algorithm}[t]
    \caption{Multi-Interest Extraction~(MIE) module}\label{alg:encode}
    \raggedright
    \textcolor{black}{\textbf{Inputs:} 
    Long-term user behavior sequence $\{s_i \in \mathbb{R}^{d \times 1}\}_{i=1}^L$, sequence length $L$, cluster number $K$, trainable parameters $W_h \in \mathbb{R}^{d \times m}$, and  distance metric function $dis(\cdot)$. \\
    \textbf{Outputs:} Interest representation $\mathbf{U} = \{u_i \in \mathbb{R}^{d \time 1}\}_{i=1}^K$.}
    \begin{algorithmic}[1]
        \Function {\textbf{MIE}}{$\{s_i\}_{i=1}^L$, $L$, $K$, $W_h$, $dis$}
            \State {\textcolor{gray}{\# Dimensionality Reduction}} 
            \For{$i \leftarrow 1$ to $L$}
                \State $h_i = W_h^T s_i$
                \State $h_p,h_n \leftarrow \text{Pos\&Neg Samples Selection}$
                \State $\alpha_i = dis(s_i, s_n) - dis(s_i, s_p)$ 
                \State $\mathcal{L}_{aux} \leftarrow max(0, dis(h_i, h_p) - dis(h_i, h_n) + \alpha_i)$
            \EndFor
            \State 
            \State {\textcolor{gray}{\# Clustering}} 
            \State $\{\mathcal{C}_i\}_{i=1}^K, \{c_i\}_{i=1}^K \leftarrow KMeans(\{h_i\}_{i=1}^L, K)$
            
            \State 
            \State {\textcolor{gray}{\# Interest Extraction}} 
            \For{$i \leftarrow 1$ to $K$}
                \State $u_i = \sum_{\forall idx \in \mathcal{C}_i} \frac{exp(sim(h_{idx}, c_i))}{\sum_{\forall j \in \mathcal{C}_i} exp(sim(h_j, c_i))} \cdot s_{idx}$ 
            \EndFor
            \State \Return  $\{u_i\}_{i=1}^K$
        \EndFunction
    \end{algorithmic}
\end{algorithm}

\subsubsection{Metric Learning based Dimensionality Reduction \label{method:dim_reduction}} 

Even for well-resourced offline platforms, existing hardware equipment still cannot meet the overhead of clustering algorithms. To reduce the computational complexity of clustering the entire sequence, we develop a novel dimensionality reduction algorithm using metric learning that can maintain the relative pairwise distances between embeddings.
Inspired by the success of ETA~\cite{chen2022efficient} and SDIM~\cite{cao2022sampling}, we begin to shift our gaze to SimHash~\cite{charikar2002similarity}. SimHash is a locality-sensitive hash~(LSH)~\cite{alexandr2015practical} algorithm, which outputs the same hash code with a high probability for nearby embeddings. Specifically, given an embedding $e \in \mathbb{R}^{d} $, SimHash calculates its binary hash code $b \in \{+1, -1\}$ by a random hash function $W_r \in \mathbb{R}^{d}$:
\begin{equation}
    b = sign(\sum_{i=1}^{d} e[i] * W_r[i]),
    \label{eq:simhash}
\end{equation}
where $W_r \sim \mathcal{N}(0, 1)$. In practice, the multi-round hash is used to improve precision, \ie, using $m$ hash functions to generate $m$-bit binary hash code. 

Motivated by this random projection-based LSH, we design an effective dimensionality reduction algorithm by introducing a random projection matrix. For simplicity, we present our algorithm in matrix form: given an embedding $e \in \mathbb{R}^{d \times 1}$, we reduce its dimension from $d$ to $m$ by a randomly initialized matrix $W_h \in \mathbb{R}^{d \times m}$:
\begin{equation}
    h = W_h^T e, 
    \label{eq:reduce_dim}
\end{equation}
where $h \in \mathbb{R}^{m \times 1}$ is the representation after dimensionality reduction of $e$, $m$ is a hyperparameter and  $W_h \sim \mathcal{N}(0, 1)$. 
By Eq.~(\ref{eq:reduce_dim}), we obtain the low-dimensionality long-term user behavior sequence representations $\mathbf{H} = \{h_i\}_{i=1}^L$ where $h_i \in \mathbb{R}^{m \times 1}$ is low-dimensionality projection of $s_i$.

The effectiveness of our dimensionality reduction algorithm heavily relies on the random projection matrix $W_h$. To keep the relative pairwise distances between user behavior embeddings consistent before and after dimensionality reduction, we take advantage of metric learning~\cite{xing2002distance} to optimize $W_h$. The objective of metric learning is to minimize the distance between same-class samples~(\ie, positive samples) while maximizing the distance between different-class samples~(\ie, negative samples) as much as possible. To this end, we design a straightforward yet impactful positive and negative sample selection strategy, along with a dynamic metric loss function.

\textit{\textbf{Positive and negative samples selection strategy.}} Given two user behaviors $h_i$ and $h_j$, it's challenging to determine whether $h_j$ is a positive or negative sample of $h_i$ since we lack relevant labels for user behaviors. However, as a dimensionality reduction algorithm, we notice that the distances between original behaviors can provide potential hints. Therefore, to obtain positive and negative samples for $h_i$~(denoted as \textit{anchor} behavior), we first randomly select two behaviors $h_j$ and $h_k$ from $ \{h_i\}_{i=1}^L$ where $ j,k \neq i $. Then we calculate the distances of original behavior embeddings~(\ie, $\{s_i\}_{i=1}^L$) between selected behaviors and the anchor behavior, which are denoted as $dis(s_i, s_j)$ and $dis(s_i, s_k)$ respectively. Here, $dis(\cdot)$ represents a distance metric function. We regard the behavior with a closer distance as positive sample $h_p$ and regard the other one as negative sample $h_n$, \ie, 
\begin{equation}
    p = \mathop{\arg\min}_{z \in \{j, k\}} \ dis(s_i, s_z), n = \mathop{\arg\max}_{z \in \{j, k\}} \ dis(s_i, s_z).
    \label{eq:pos_neg}
\end{equation}

\textit{\textbf{Dynamic triplets loss.}} After obtaining positive and negative samples, we utilize the triplets loss~\cite{schroff2015facenet} to optimize $W_h$. The triplets loss $\mathcal{L}_{aux}$ is formulated for each sample as:
\begin{equation}
    \mathcal{L}_{aux} = \sum_{i=1}^L{max(0, dis(h_i, h_p) - dis(h_i, h_n) + \alpha)},
    \label{eq:loss}
\end{equation}
where $\alpha$ is a hyperparameter. The triplets loss can pull the difference between $dis(s_i, s_n)$ and $dis(s_i, s_p)$ to larger than the margin $\alpha$, \ie, $dis(h_i, h_n) - dis(h_i, h_p) \geq \alpha$. One potential problem is that if the sampled two behaviors are very close (\ie, the ground truth value of $dis(h_i, h_n) - dis(h_i, h_p)$ is close to 0) the adopted triplets loss will mislead the optimize direction. To address this concern, we again utilize the original behavior embeddings and calculate the value of $\alpha_i$ dynamically for each anchor behavior $h_i$ by:
\begin{equation}
     \alpha_{i} = dis(s_i, s_n) - dis(s_i, s_p).
     \label{eq:alpha}
\end{equation}

In this way, even if the sampled positive and negative behaviors are very close, our dynamic triplets loss still generates a right supervised signal, \ie, margin $\alpha_i$ will be close to 0. 

\subsubsection{Clustering Module} By clustering, we can divide the user's historical behaviors into multiple clusters of similar behaviors in an unsupervised manner. Therefore, after getting low-dimension user behavior embeddings $\mathbf{H} = \{h_i\}_{i=1}^L$, we use a clustering module to obtain $K$ clusters $\mathbf{C} = \{\mathcal{C}_i\}_{i=1}^K$ and corresponding clustering center $\mathbf{c} = \{c_i\}_{i=1}^K$ where $K$ is pre-defined cluster number and $\mathcal{C}_i$ is a set of user behavior indexes belonging to $i$-th cluster. Moreover, we adopt the same metric function~(\eg, $dis(\cdot)$) to measure the distance between behaviors when clustering, which helps to satisfy \textbf{R2}. 
In our experiments, we use KMeans to accomplish the clustering process due to its fast convergence.

\subsubsection{Offline Multi-Interest Extraction} Then, we present a method for extracting multi-interest representations of user clusters. Conventionally, clusters are characterized by their centers. However, we contend that such centers obtained through auto-clustering are only approximate representations of the respective clusters, lacking in accuracy. For instance, in the case of KMeans clustering, the center is merely an average depiction of all the elements in the cluster, akin to an average pooling layer.
To obtain a more accurate cluster representation $\mathbf{U}=\{u_i\}_{i=1}^K$, we apply the target attention mechanism~\cite{zhou2018deep} within clusters, taking the cluster center as the query:

\begin{equation}
    u_i = \sum_{\forall idx \in \mathcal{C}_i} \frac{exp(sim(h_{idx}, c_i))}{\sum_{\forall j \in \mathcal{C}_i} exp(sim(h_j, c_i))} \cdot s_{idx}, 
    \label{eq:offline_ta}
\end{equation}
\begin{equation}
    sim(h_i, h_j) = \frac{1 - dis(h_i, h_j)}{\beta},
    \label{eq:sim}
\end{equation}
where $dis(\cdot)$ is the distance metric function and $\beta$ is the scaling factor. To avoid missing \textbf{R2}, we utilize $sim(\cdot)$ as a similarity metric function which is based on our distance metric function, rather than the scaled dot-product used in standard target attention. Thus, user behaviors that are close to the cluster center will get greater weights. 

\subsection{Online Inference\label{sec:online}} 
To obtain the user's interest representation relative to target items for online inference, we again use target attention to model the pre-processed offline multi-interest representations $\mathbf{U}=\{u_i\}_{i=1}^K$:
\begin{equation}
    \mathbf{I} = \sum_{i=1}^K \frac{exp(sim(u_i, x_t))}{\sum_{j=1}^K exp(sim(u_j, x_t))} \cdot u_i, 
\end{equation}
where $x_t$ is target item and $\mathbf{I}$ is user's interest representation relative to target item. 
Clearly, both online calculation overhead and performance are related to $K$. 
In the extreme case where $K$ equals the length of behavior sequence $N$, our ENCODE is equivalent to target attention, achieving optimal performance but the worst efficiency. Conversely, if $K$ equals 1, ENCODE almost brings no extra overhead, \ie, outputs an off-the-shelf interest representation.
By adjusting $K$, we can achieve an optimal balance between performance and online efficiency~(\eg, $K$ is set to 30 empirically).

\begin{table*}[t]
    \normalsize
    \begin{center}
    \caption{Time complexity of different methods during the online-inference stage. The ``\textbf{\textit{Len}(Seq.)}" represents the utilized behavior sequence length. $B$ is the number of candidate items. $L$ is the length of the original behavior sequence. $k$ is the retrieved length for online search-based methods. $A$ is the size of the attribute inverted index in SIM. $n$ is the number of hashes. $m$ is the dimensionality of our low-dimensional embeddings. $K$ and $T$ are the numbers of clusters and iterations of KMeans in our ENCODE. In our experiments on industrial dataset, $A=1$, $k=50$, $L=1000$, $n=64$, $d=32$, $m=4$, $T=15$, $K=30$ and $B \approx 3000$.}
    
    \label{tab:complexity}
        \begin{tabular}{l | l  l c}
            \toprule[1pt]
                \textbf{Method} & \textbf{Online-Inference} & \textbf{Offline-Training} & \textbf{\textit{Len}(Seq.)}\\
            \midrule[0.5pt]
                DIN~\cite{zhou2018deep} & $O(BLd)$ & $O(BLd)$ & $L$\\
                SIM~\cite{pi2020search} & $O(B\log A + Bkd)$ & $O(B\log A + Bkd)$ & $k$ \\
                ETA~\cite{chen2022efficient} & $O(BLn + Bkd)$ & $O(Ln \log d + BLn + Bkd)$ & $k$ \\ 
                SDIM~\cite{cao2022sampling} & $ O(Bn\log d)$ & $ O(Ln \log d + Bn\log d)$  &  $L$\\
                ENCODE & $O(BKd)$ & $O(Ldm+TKLm+Lm+BKd)$ &  $L$ \\
            \bottomrule[1pt]
        \end{tabular}
    \end{center}
    \vspace{-2em}
\end{table*}

\subsection{Complexity Analysis}

In this subsection, we analyze the time complexity of ENCODE of offline and online stages. 

\textbf{\textit{Offline Extraction Stage.}} We analyze the time complexity step by step. \textit{(1) Dimensionality reduction.} At this step, we reduce the user's $L$ historical behaviors from $d-$dimension to $m-$dimension, so the time complexity is $O(Ldm)$. \textit{(2) Clustering.} In practice, we utilize KMeans to accomplish this step. Specifically, KMeans calculates the distances between $L$ historical behaviors and $K$ cluster centers at each iteration, and the maximum number of iterations is $T$. Thus, the time complexity of clustering is $O(TKLm)$. \textit{(3) Multi-interest extraction.} For each historical behavior, ENCODE obtains its attention weight by calculating the similarity score with its corresponding cluster center. Therefore, the time complexity is $O(Lm)$. Thus, the total time complexity of the offline stage is $O(Ldm+TKLm+Lm)$. 

\textbf{\textit{Online Inference Stage.}} Suppose the number of candidate items is $B$, and for every candidate item, ENCODE obtains attention weights for $K$ interest representations by calculating similarity scores between interest representations and the target item, then the time complexity of online stage is $O(BKd)$.

\textbf{\textit{Advantages.}} Table~\ref{tab:complexity} demonstrates the time complexity comparison between our ENCODE and other methods for long-term sequence modeling, both during the online inference stage and the offline training stage.
We begin with the time complexity of the online inference stage. In comparison to online search-based methods~(\ie, SIM and ETA), our ENCODE method utilizes the entire sequence while having a comparable time complexity. Additionally, when compared to sampling-based methods~(\ie, SDIM), our ENCODE method surpasses it in performance by unifying relevance metrics~(\ie, satisfy \textbf{R2}), despite being slightly less efficient. Overall, our ENCODE algorithm achieves a comparable performance to target attention~(\eg, DIN), while the extra overhead brought by ENCODE is acceptable, \ie, equivalent to modeling an additional sequence of length $K$. 
Moving on to the offline training stage, although we delegate the complex interest extraction module to the offline stage, our designed dimensionality reduction algorithm ensures that our offline training has a similar complexity to SIM.

\section{Experiments}

\subsection{Experimental Settings}
\textbf{Dataset.} We evaluated ENCODE on both the public and industrial datasets. 
\revise{For the public dataset, we used the Amazon Books~\cite{julian2015image,lin2022sparse} and MovieLens 32M~\cite{harper2015movielens}. Specifically, Amazon Books comprises 295,982 users, 647,578 items, and 6,626,872 samples, and MovieLens 32M includes 200,948 users, 87,585 items, and 32,000,204 samples. We split the samples into 80\% training and 20\% test data according to the behavior timestamp. The most recent 50 behaviors are selected as the real-time sequence while the recent 1,500 behaviors are selected as the long-term sequence.}
The industrial dataset is collected from one of the largest international e-commerce platforms (\ie AliExpress.com), involving a billion scales of users and items. We selected consecutive 14 days of data for training and the next day's data for testing. The industrial dataset consists of over 836 million samples, which contain both positive~(\ie, exposure with user clicks) and negative~(\ie, exposure without user clicks) samples. Specifically, we regarded the recent 50 behaviors as the user's real-time sequence and regarded the recent 1,000 behaviors as the user's long-term behavior sequence. 

\textbf{Evaluation Metrics.} For offline experiments, we followed previous work~\cite{cao2022sampling,chen2022efficient} and reported the \textbf{Area Under Curve~(AUC)} indicator, which reflects the ability of the CTR model to rank positive samples in front of negative samples. Besides, we also reported the \textbf{Group Area Under Curve~(GAUC)} indicator. Specifically, we calculated GAUC at the page view level to reflect ranking ability within a page. For the \revise{public} datasets, we only reported \textbf{AUC} since we lack relevant group information. 
Additionally, we used the \textbf{Inference Time} as a supplement metric to show the efficiency of each model by deploying models online to serve user requests that are copied from the product environment. The machines and the number of user requests are controlled the same for fairness comparison.

\subsection{Implementation Details}
\textbf{ENCODE Settings.} 
For dimensionality reduction, the original dimensionality of user behavior embeddings is $32$ for the industrial dataset and public dataset, and the target low-dimensionality $m$ was set to $4$. For the clustering module, we used KMeans, and the cluster number $K$ was set to $30$ for both industrial and public datasets. For distance metric function $dis(\cdot)$ throughout the online and offline stages, we utilized cosine similarity to measure the distance between user behaviors, \ie, 
\begin{equation}
    dis(s_i, s_j) = 1 - \frac{s_i \cdot s_j}{\parallel s_i\parallel_2 \cdot \parallel s_j\parallel_2}.
\end{equation}
For the similarity metric function in the target attention module, we set the scaling factor $\beta$ as $20$.

\textbf{Training Details.} Following industry practice to prevent over-fitting, we trained all methods only for the $1$ epoch. We used the Adam~\cite{kingma2014adam} algorithm as the optimizer with a learning rate $1e-4$ and batch size $1024$ for all experiments. All model parameters were initialized by random. Loss weights of $\mathcal{L}_{CE}$ and $\mathcal{L}_{aux}$ were set to $1$ and $0.1$.

\begin{table*}[t]
    \begin{center}
    \small
    \caption[]{Performance comparison with SOTA methods on industrial and public datasets. We report the \textit{mean} and \textit{std} over 3 replicate experiments, \revise{\ie, mean $\pm$ std}. \revise{\textbf{``Inference Time"} represents the average serving time for each request when deployed online.} The best and second-best results are highlighted in bold and underlined respectively.} 
    \label{tab:cmp_with_sota}
        \begin{tabular}{ l |  c  c  c   | c | c }
            \toprule[1pt]
                \multirow{2}{*}{\textbf{Methods}}
                 & \multicolumn{3}{c|}{\textbf{Industrial Dataset}} & \textbf{\revise{Amazon Books}} & \textbf{\revise{MovieLens 32M}} \\
               & \textbf{CTR AUC} & \textbf{CTR GAUC} & \textbf{Inference Time} & \textbf{CTR AUC} & \textbf{\revise{CTR AUC}}  \\  
            
            \midrule[0.5pt]
                 DIN~\cite{zhou2018deep} & 0.71739 \scriptsize{$\pm~0.00017$} & 0.67864 \scriptsize{$\pm~0.00032$ } & 11.5 ms &  0.63573 \scriptsize{$\pm~0.00057$} & \revise{0.79889 \scriptsize{$\pm~0.00053$}}\\
                
                AvgPooling & 0.71985 \scriptsize{$\pm~0.00039$} & 0.68180 \scriptsize {$\pm~0.00047$} & 12.5 ms &   0.86151 \scriptsize{$\pm~0.00082$} & \revise{0.80248 \scriptsize{$\pm~0.00046$}}\\
                
                DIN-L~\textit{(upper bound)} & \underline{0.72290 \scriptsize{$\pm~0.00018$}} & \textbf{0.68651} \scriptsize{$\pm~0.00007 $} & 26.7 ms &  \underline{0.88281 \scriptsize{$\pm~0.00037 $}} & \revise{\textbf{0.80778} \scriptsize{$\pm~0.00054$}}\\
                
                 SIM (Hard)~\cite{pi2020search} & 0.72206 \scriptsize{$\pm~0.00009 $} & 0.68613 \scriptsize{$\pm~0.00003 $} & 14.4 ms &  0.88077 \scriptsize{$\pm~0.00071$} & \revise{0.80359 \scriptsize{$\pm~0.00017$}}\\
                 
                 UBR4CTR~\cite{Qin0WJF020} & 0.72182 \scriptsize{$\pm~0.00023 $} & 0.68623 \scriptsize{$\pm~0.00013 $} & 19.2 ms &  0.87715 \scriptsize{$\pm~0.00061$} & \revise{0.80513 \scriptsize{$\pm~0.00039$}}\\

                 ETA~\cite{chen2022efficient} & 0.72181 \scriptsize{$\pm~0.00007 $} & 0.68587 \scriptsize{$\pm~0.00047 $} & 16.2 ms &  0.88163 \scriptsize{$\pm~0.00034$} & \revise{0.80566 \scriptsize{$\pm~0.00038$}}\\
                  
                  TWIN~\cite{chang2023twin} &  0.72224 \scriptsize{$\pm~0.00011 $} & 0.68590 \scriptsize{$\pm~0.00014 $} & 18.0 ms & 0.87474 \scriptsize{$\pm~0.00055 $} & \revise{0.80536 \scriptsize{$\pm~0.00087$}}
                  \\ 
                  
                SDIM~\cite{cao2022sampling} & 0.72210 \scriptsize{$\pm~0.00014$} & 0.68551 \scriptsize{$\pm~0.00012$}  & 13.1 ms &  0.86564 \scriptsize{$\pm~0.00042$} & \revise{0.80649 \scriptsize{$\pm~0.00075$}}\\
            \midrule[0.5pt]     
                ENCODE (Ours) & \textbf{0.72312} \scriptsize{$\pm~0.00022$} & \underline{0.68626 \scriptsize{$\pm~0.00010$}}  & 13.6 ms &  \textbf{0.88402} \scriptsize{$\pm~0.00041$} & \revise{\underline{0.80772 \scriptsize{$\pm~0.00031$}}} \\
            \bottomrule[1pt]
        \end{tabular}
    \end{center}

\end{table*}

\subsection{Competitive Models\label{sec:sota_methods}}

To demonstrate the effectiveness of our proposed ENCODE, we compared it with the following state-of-the-art~(SOTA) methods of modeling user's long-term behavior sequence:

\begin{itemize} [left=1em]
    \item \textbf{DIN~\cite{zhou2018deep}.} DIN is the most widely-used method to model users' short-term behavior sequence by target attention~(TA). In our experiments, DIN is regarded as a baseline without utilizing long-term behavior sequences.
    \item \textbf{AvgPooling.} Based on DIN, AvgPooling uses an average pooling operation to model the user's long-term sequence.
    \item \textbf{DIN-L.} DIN-L is an extension of DIN, which leverages TA to model the user's long-term behavior sequence. But unfortunately, it cannot be deployed online. We treat the performance of DIN-L as the upper bound.
    \item \textbf{MIMN~\cite{ren2019lifelong}.} MIMN is a typical memory-based method that designs a separate module to maintain the user's latest interests. When serving online, MIMN takes off-the-shelf interests to obtain long-term interest with an RNN-based network. 
    \item \textbf{{SIM (Hard)~\cite{pi2020search}.}} SIM (Hard) is a search-based method that retrieves top-$k$ most relevant behaviors with target item by category and then models them by TA. 
    \item \textbf{UBR4CTR~\cite{Qin0WJF020}.} Different from the retrieval process of SIM~(Hard), UBR4CTR designs a feature selection module to generate the query for retrieval. According to the query, $k$ historical behaviors are retrieved and modeled by TA. 
    \item \textbf{ETA~\cite{chen2022efficient}.} ETA adopts LSH~(\ie, SimHash) algorithm to generate binary hash codes for the target item and historical behaviors, and then retrieves $k$ historical behaviors by hamming distance for subsequent TA.
    \item \textbf{TWIN~\cite{chang2023twin}.} TWIN is another search-based method that ensures consistency throughout different stages by leveraging the scaled dot-product~(\ie, used in TA) for retrieving.
    \item \textbf{SDIM~\cite{cao2022sampling}.} SDIM is a sampling-based method that also adopts the SimHash algorithm, then samples behaviors sharing the same signatures with the target item. 
\end{itemize}

We reproduced each CTR model using their respective configurations and conducted a full investigation of hyperparameters on the used dataset, selecting the best ones. 
Following~\cite{cao2022sampling}, we omitted the baseline \textbf{MIMN~\cite{ren2019lifelong}}, since MIMN and SIM are proposed by the same team and the authors, claimed that SIM defeats MIMN. 
As the Amazon dataset lacks item side information, for SIM, we retrieved top-$k$ behaviors by item ID, and for UBR4CTR, we used a strengthened version of the feature selection module with multi-head self-attention. For a fair comparison, we used the same input features and model architecture except for the long-term user behavior sequence modeling module.

\subsection{Offline Evaluation}
To demonstrate the effectiveness of our ENCODE, we compared it with SOTA methods on both industrial and public datasets. For a clearer comparison, we group existing methods into 3 parts: (1) Baselines: \textbf{DIN~\cite{zhou2018deep}}, \textbf{AvgPooling} and \textbf{DIN-L}. (2) Online search-based Methods: \textbf{SIM (Hard)~\cite{pi2020search}}, \textbf{UBR4CTR~\cite{Qin0WJF020}}, \textbf{ETA~\cite{chen2022efficient}} and \textbf{TWIN~\cite{chang2023twin}}. (3) Offline Interest Extraction based Methods: \textbf{SDIM~\cite{cao2022sampling}} and our proposed \textbf{ENCODE}. All results are reported in Table~\ref{tab:cmp_with_sota}. 

\subsubsection{Results on Industrial Dataset} From Table~\ref{tab:cmp_with_sota}, several observations can be made. \textbf{(1)} All methods that introduce long-term sequences perform significantly better than DIN~\cite{zhou2018deep}\footnote{With over 836 million samples in our industrial dataset, even a slight $0.05\%$ improvement in AUC and GAUC during offline evaluation can lead to significant online business gains.}. This includes the naive AvgPooling method, demonstrating that modeling the user's long-term sequence is effective. \textbf{(2)} Offline methods are more efficient and outperform almost all online search-based methods, proving that modeling the entire long-term behavior sequence without discarding any behaviors (i.e. satisfying \textbf{R1}) improves performance. \textbf{(3)} Our proposed ENCODE achieves the best performance, almost reaching the upper bound of DIN-L, while its inference time is only about half that of DIN-L.
Additionally, we observe that the performance of SIM~(Hard) is slightly superior to soft search ETA. This is attributed to the fact that our industrial dataset encompasses a wider range of user behaviors, and SIM employs a leaf category for retrieval, which guarantees the relevance of the retrieved behaviors.



\subsubsection{Results on Public Dataset} 

\revise{
According to the results presented in Table.~\ref{tab:cmp_with_sota}, we obtained similar results on public datasets: \textbf{(1)} Long-term sequence modeling yields significant improvements, \eg, achieving over a 22.58\% gain in CTR AUC on Amazon Books. \textbf{(2)} Our proposed ENCODE outperforms all competitive methods and performs comparably to DIN-L. \textbf{(3)} Interestingly, another offline method SDIM~\cite{cao2022sampling} shows good performance on MovieLens 32M but fails to surpass search-based methods on Amazon Books. We attribute this phenomenon to the gap between hash collision probabilities and attention weights, resulting from not meeting \textbf{R2}.
}

\subsection{Hyper-Parameter Analysis}

\begin{figure}[t]
    \centering
    \vspace{-2em}
    \subfigure[]{
        \includegraphics[width=0.94\linewidth]{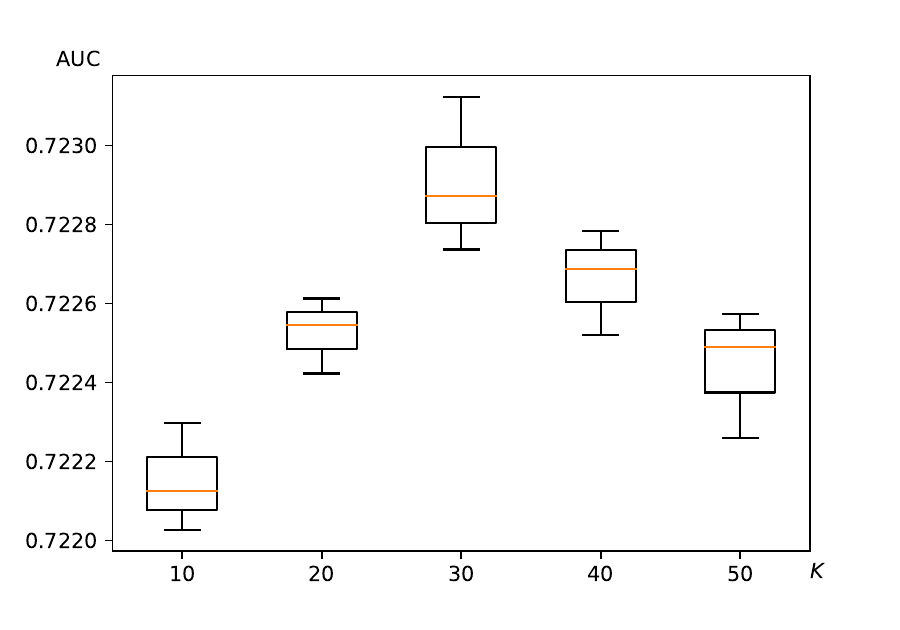}
    } 
    \subfigure[]{
        \includegraphics[width=0.94\linewidth]{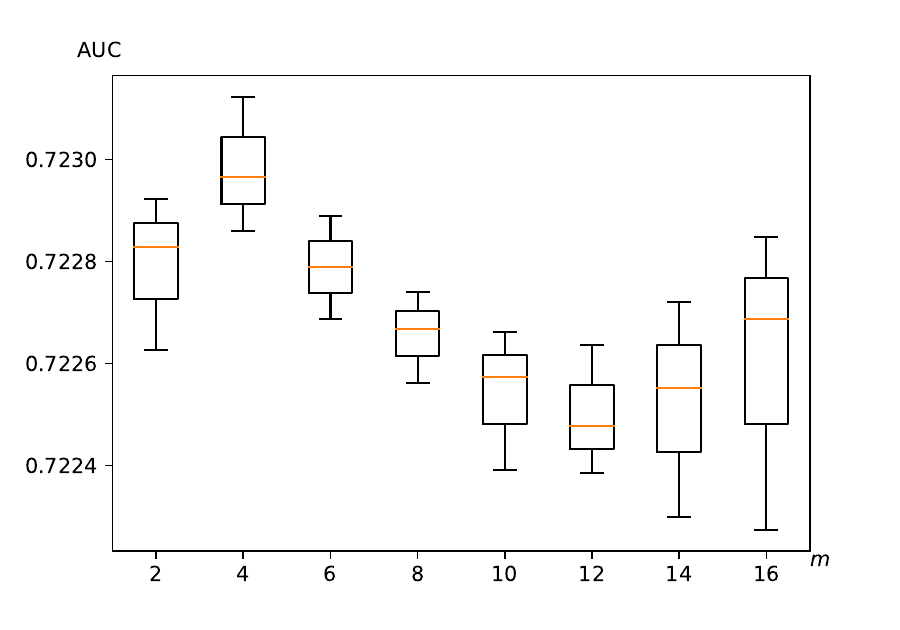}
    }
    \caption{Performance on the industrial dataset of different hyperparameters of ENCODE. (a) The results under varied cluster number $K$. (b) The results under varied $m$.}
    \label{fig:influece_mk}
\end{figure}

We conducted several experiments to analyze the influence of different hyperparameters of ENCODE, including cluster number $K$ and the target dimensionality $m$ in dimensionality reduction. \revise{The results are presented as box plots in Fig.~\ref{fig:influece_mk}}.

\textbf{Influence of Cluster Number $K$.} We evaluated our ENCODE using various values of $K$, where $K \in \{10, 20, 30, 40, 50\}$. We also attempted to use a larger value of $K$~(\eg, $K=100$) but encountered an out-of-memory error during offline CTR model training. As shown in Fig. ~\ref{fig:influece_mk}(a), our ENCODE achieves the best CTR AUC and GAUC when $K=30$. Notably, we observe that the performance decreased when $K$ exceeds $30$, possibly due to an increased cluster number resulting in insufficient behaviors per cluster to accurately represent user interests.

\textbf{Influence of Target Dimensionality $m$.} \revise{
We compared the performance of different values of $m$, where $m \in \{2, 4, 6, 8, 10, 12, 14, 16\}$. As shown in Fig.~\ref{fig:influece_mk}(b), we observe that as $m$ increases, the performance of ENCODE improves, peaking at $m=4$. Beyond $m=4$, the performance initially decreases before improving again. We believe this phenomenon arises because a larger $m$ allows the dimensionality reduction module to retain more details, but it also retains some noise~(since the embedding before dimensionality reduction is trained from scratch and is not perfect), thereby affecting the model's generalization. Another piece of supporting evidence is that with a larger $m$, the variance of the training results also increases. 
}

\subsection{Ablation Studies}

To validate the effectiveness of ENCODE, we conducted extensive ablation studies by answering the following research questions~(\textbf{RQ}): 
\begin{itemize} [left=1em]
    \item \textbf{RQ1:} Does ENCODE extract more relevant interest representations to the target item \ie, satisfying \textbf{R2}? 
    \item \textbf{RQ2:} Does ENCODE gain improvements from utilizing a consistent relevance metric function?
    \item \textbf{RQ3:} Does our dimensionality reduction algorithm generate a more accurate low-dimensionality representation than SimHash?  
    \item \textbf{RQ4:} What's the influence of positive and negative sample selection strategy in metric learning?
    \item \textbf{RQ5:} What's the influence of loss function in metric learning?
    \item \textbf{RQ6:} Can ENCODE be extended to other clustering algorithms? 
\end{itemize}

\begin{table}[t]
    \small
    \begin{center}
    \caption[]{Performance on the industrial dataset of different relevance metric functions in TA. $\mathcal{RI}$ denotes the Relevance Indicator, which is used to measure the relevance between the long-term interest representations and target items. The best and second-best results are in bold and underlined.}
    \label{tab:ablation_relevance}
        \begin{tabular}{ l |  c  c | c }
        \toprule[1pt]
            \multirow{2}{*}{\textbf{Methods}} & \multicolumn{2}{c|}{\textbf{CTR}} & \multirow{2}{*}{$\mathcal{RI}$ $\downarrow$}\\  
            &  \textbf{AUC} & \textbf{GAUC} & \\
            \midrule[0.5pt]
            DIN~\cite{zhou2018deep} &  0.71739 & 0.67864 & --- \\
            DIN-L~\textit{(upper bound)} & \underline{0.72290} & \underline{0.68651} & --- \\
            SIM (Hard)~\cite{pi2020search} & 0.72206  & 0.68613  & 6.910 \\
                 
            UBR4CTR~\cite{Qin0WJF020} & 0.72182 & 0.68623 & 6.932 \\
                 
            ETA~\cite{chen2022efficient} & 0.72181 & 0.68587  & 7.032 \\

            TWIN~\cite{chang2023twin} & 0.72224 & 0.68590 & 6.759 \\
            
            SDIM~\cite{cao2022sampling} & 0.72210 & 0.68551 & 6.919 \\
        \midrule[0.5pt]
            ENCODE$^-$ (Ours) &  0.72250 & \textbf{0.68663} & \underline{6.731} \\
            ENCODE (Ours) &  \textbf{0.72312} & 0.68626 & \textbf{6.170}\\
        \bottomrule[1pt]
    \end{tabular}
    \end{center}  
\end{table}

\subsubsection{Relevance Between Extracted Interests and the Target Items} 
For the relevance between the extracted long-term interest representations and target item embeddings, it is difficult to measure directly due to their differing feature spaces. 
However, from another perspective, the weights assigned by the model to historical behaviors during the modeling of long-term interest can serve as an indicator of whether the model successfully extracts interest from the most relevant behaviors. In other words, if we have an ideal weights distribution that assigns greater weights to more relevant historical behaviors, we can compare the relevance between the extracted long-term interest representations and target item embeddings by examining the difference between the model's weights and the ideal weights.
Considering that \textbf{DIN-L} is recognized as the advanced accurate long-term sequence modeling algorithm, we regard the attention weights of DIN-L as ground truth and define \textbf{Relevance Indicator~($\mathcal{RI}$)} as the distance between the output weights of each method and attention weights of DIN-L, to measure the relevance. 
Specifically, we use cross-entropy as the distance metric. For ENCODE, we use the product of within-cluster and between-cluster weights as final weights.
We compared the $\mathcal{RI}$ for all methods. All results are reported in Table~\ref{tab:ablation_relevance}.

From the results, it is evident that our proposed ENCODE surpasses all SOTA methods in terms of the $\mathcal{RI}$. These considerable improvements serve as evidence that our ENCODE effectively extracts more relevant long-term interest representations to the target items. Furthermore, we observe a correlation between models that achieve better CTR performance and better $\mathcal{RI}$. This observation further supports the notion that meeting \textbf{R2} indeed contributes to better overall performance.

\subsubsection{Effect of Unifying Relevance Metric Function~\textbf{(RQ2)}} To show the effectiveness of unifying relevance metric function, we tried two different relevance metric functions in ENCODE's TA module: \textbf{\textit{(1)}} Scaled cosine similarity, \ie, Eq.~(\ref{eq:sim}). \textbf{\textit{(2)}} Scaled dot-product used in standard TA, denoted \textbf{ENCODE$^-$}. The latter suffers from inconsistent relevance metric function issues. All results are reported in Table~\ref{tab:ablation_relevance}.

From Table~\ref{tab:ablation_relevance}, it is evident that by unifying the relevance metric function, ENCODE achieves a 0.062\% improvement in CTR AUC and demonstrates better performance on $\mathcal{RI}$, which indicates that ENCODE can extract more relevant interest representations for the target item.
Further evidence supporting the effectiveness of unifying the relevance metric function is TWIN. In comparison to other search-based methods, TWIN utilizes the same relevance metric function across different steps and achieves the best performance during search-based methods. But unfortunately, the limited retrieval length~(\ie, not satisfy \textbf{R1}) is the bottleneck of its performance.
In addition, even taking scaled dot-product as TA's relevance metric function, our proposed ENCODE still outperforms all SOTA methods on CTR AUC and $\mathcal{RI}$, underscoring the effectiveness of our ENCODE.


\begin{table}[t]
    \small
    \begin{center}
    \caption[]{Performance on the industrial dataset of different variants of ETA. ``$m$" represents the embedding dimension participating in the distance measure, \eg, the bit-length of hashcode for ETA. For standard ETA, we followed the official setting and set the bit-length of the hashcode as 2 times the embedding size, \ie, $m = 64$. The best and second-best results are in bold and underlined.}
    \label{tab:cmp_with_simhash}
        \begin{tabular}{ l | c |  c  c }
            \toprule[1pt]
            \textbf{Methods} & $m$ & \textbf{CTR AUC} & \textbf{CTR GAUC} \\ 
            \midrule[0.5pt]
                DIN~\cite{zhou2018deep} & -& 0.71739 & 0.67864 \\
            \midrule[0.5pt]
                ETA~\cite{chen2022efficient} & 64 & 0.72181 & \underline{0.68587} \\
                ETA-ENCODE & 4 &  \textbf{0.72242} & \textbf{0.68607} \\
                ETA-TA & 32 & \underline{0.72230} & 0.68568  \\
            \bottomrule[1pt]
        \end{tabular}
    \end{center}
\end{table}

\subsubsection{Our Rimensionality Reduction vs. SimHash~\textbf{(RQ3)}} To answer \textbf{RQ3}, we adopted the architecture of ETA~\cite{chen2022efficient}. ETA leverages SimHash~\cite{charikar2002similarity} to encode user behavior into integer binary hash code, then retrieves top-$k$ most relevant behaviors by hamming distance. For fairness comparison, we only replaced the relevance measure algorithm of ETA, and proposed two variants of ETA: \textbf{(1) ETA-ENCODE.} ETA-ENCODE accomplishes retrieval by taking our dimensionality reduction and distance metric function $dis(\cdot)$. \textbf{(2) ETA-TA.} ETA-TA directly measures the relevance by relevance metric function in standard TA~(\ie, scaled dot product) between \textbf{original} user behaviors and target item. Both ETA-ENCODE and ETA-TA retrieve the same number of behaviors as ETA~(\eg, $k=50$). All results are reported in Table~\ref{tab:cmp_with_simhash}. 

From the results, we can observe that ETA-ENCODE surpasses ETA by an obvious margin~(\eg, 0.061\% on CTR), even if ETA adopts $64$ hash functions while our reduced embeddings are 4-dimensional. ETA-ENCODE achieves comparable performance with ETA-TA, which demonstrates our dimensionality reduction algorithm effectively maintains the relative pairwise distances between behaviors.

\begin{table}[t]
    \begin{center}
    \small
    \caption[]{Performance on the industrial dataset of different positive and negative sample selection strategies. \textbf{``Pos\&Neg Selection"} denotes different pos\&neg sample selection strategies. The best and second-best results are in bold and underlined.}
    \label{tab:ablation_selection}
    \begin{tabular}{ l |  c |  c  c }
        \toprule[1pt]
            \multirow{2}{*}{\textbf{Methods}} & \multirow{2}{*}{\textbf{Pos/Neg Selection}} & \multicolumn{2}{c}{\textbf{CTR}} \\ 
            &  & \textbf{AUC} & \textbf{GAUC} \\
            \midrule[0.5pt]
            DIN~\cite{zhou2018deep} & --- & 0.71739 & 0.67864 \\
        \midrule[0.5pt]
            \multirow{3}{*}{ENCODE}
                & Within Neighbors & 0.72243  & \textbf{0.68664} \\
                & Within Sequence & \textbf{0.72312 } & \underline{0.68626} \\
                & Within Batch &  \underline{0.72273} & {0.68594}  \\
        \bottomrule[1pt]
    \end{tabular}
    \end{center}  
\end{table}

\subsubsection{Influence of Samples Selection Strategy~\textbf{(RQ4)}\label{sec:ab_posneg}} To answer \textbf{RQ4}, we first compared different pos/neg samples strategy. To prevent the sampling strategy from introducing additional computational overhead, we kept random sampling but adopted different sampling ranges: \textbf{(1) ``Within Neighbors: "} pos\&neg samples came from previous or next behavior in sequence. \textbf{(2)``Within sequence:"} pos\&neg samples came from the same behavior sequence. \textbf{(3) ``Within Batch: "} pos\&neg samples came from whole batch. The results are reported in Table~\ref{tab:ablation_selection}. 


Based on the obtained results, it is evident that our ENCODE consistently outperforms DIN by a significant margin when employing different sample selection strategies. This robust performance proves the effectiveness of our dynamic triplets loss across various sampling strategies. Remarkably, the sampling within sequence strategy achieves the best CTR performance~(\ie, 0.72312).
Additionally, we observed that sampling within neighbors yields the worst performance. Upon analysis, we attribute this to the locality of user behaviors, wherein adjacent behaviors often tend to be very similar or even the same\footnote{In fact, approximately 2.8\% of historical behaviors in our industrial dataset are adjacent to the same behavior.}. This similarity among adjacent behaviors poses challenges for optimization and limits the diversity of selected samples.

\begin{table}[t]
    \small
    \begin{center}
        \caption[]{Performance on the industrial dataset of different relevance metric functions in TA. The best and second-best results are in bold and underlined.}
        \label{tab:ablation_loss}
        \begin{tabular}{ l |  c |  c  c }
            \toprule[1pt]
            \multirow{2}{*}{\textbf{Methods}} & \multirow{2}{*}{\textbf{Loss Function}} & \multicolumn{2}{c}{\textbf{CTR}} \\ 
            &  & \textbf{AUC} & \textbf{GAUC} \\
            \midrule[0.5pt]
                DIN~\cite{zhou2018deep} & --- & 0.71739 & 0.67864 \\
                AvgPooling & --- & 0.71985 & 0.68180 \\
                DIN-L~\textit{(upper bound)} & --- & \underline{0.72290} & \textbf{0.68651}\\
            \midrule[0.5pt]
                \multirow{5}{*}{ENCODE}
                & None & 0.72209 & 0.68614 \\
                & MSE & 0.72021 & 0.68270 \\
                & N-Pair-MC & 0.72218 & 0.68557 \\
                & Triplets-Fixed & 0.72255 & 0.68593 \\
                & Triplets-Dynamic & \textbf{0.72312} & \underline{ 0.68626}\\
            \bottomrule[1pt]
        \end{tabular}
    \end{center}  
\end{table}

\subsubsection{Influence of Loss Function~\textbf{(RQ5)}} To answer \textbf{RQ5} and validate the effectiveness of our proposed dynamic triplets loss, we compared different loss functions for metric learning: \textbf{(1) ``None":} without any loss functions. \textbf{(2) ``MSE": } mean squared error, \ie, $\mathcal{L}_{mse} = \big(dis(h_i, h_j) - dis(s_i, s_j)\big)^2$,
where $j \in [1, L] $ is a random index. \textbf{(3) ``N-Pair-MC"}: proposed by \cite{sohn2016improved} which is extended to multiple negative samples, \ie, 
\begin{equation}
    \mathcal{L}_{N-Pair-MC} = log \big(1 + \sum_n exp(h_i^T \cdot h_n - h_i^T \cdot h_p)\big),
\end{equation} 
where $n, p \in [1, L] $ is the index of negative and positive behavior. Similar to our pos\&neg sampling strategy, we sampled negative and positive behaviors within the sequence and decided positive sample by original behavior embedding. In this experiment, the number of negative samples is set to $5$. \textbf{(4) ``Triplets-Fixed": } original triplets loss with a fixed margin $\alpha$, \eg, $\alpha=0.2$. \textbf{(5) ``Triplets-Dynamic": } our proposed dynamic triplets loss function.


From the results presented in Table~\ref{tab:ablation_loss}, it can be observed that despite ENCODE utilizing random projections for dimensionality reduction~(\ie, without loss function), its performance still surpasses AvgPooling~(\eg, 0.72209 vs. 0.71985 on CTR AUC). Compared to fixed-margin triplets loss, our proposed dynamic triplets loss achieves a 0.057\% improvement in CTR AUC and a 0.033\% improvement in CTR GAUC, demonstrating the effectiveness of the dynamic margin. Additionally, the performance of the dynamic triplets loss is superior to all the aforementioned loss functions, even when \textbf{``N-Pair-MC"} incorporates multiple negative samples. Moreover, we notice that if ENCODE were to use MSE as the loss function, the performance would be even worse than without a loss function~(\eg, 0.72021 vs. 0.72209 on CTR AUC). This can be attributed to the fact that MSE aims to ensure complete consistency in the distances between behaviors before and after dimensionality reduction, which is a highly stringent learning objective and difficult to optimize, resulting in subpar performance. Conversely, our proposed dynamic triplets loss only focuses on maintaining consistent relative pairwise distances between behaviors, which is relatively easier to optimize.

\begin{table}[t]
    \begin{center}
    \small
    \caption[]{\revise{The performance and time complexity of different clustering methods on the industrial dataset. \textbf{``Dy. Routing"} and \textbf{``AGG."} represents ``Dynamic Routing" and ``Agglomerative" clustering algorithm. The best and second-best results are highlighted in bold and underlined. For symbols used in time complexity, $L$ denotes the length of the behavior sequence, $m$ is the dimensionality of low-dimensional embeddings, $K$ is the cluster number, and $T$ is iterations of KMeans and dynamic routing. In our experimental settings, $L=1000$, $m=4$, $T=15$, $K=30$. }}
    \label{tab:ablation_cluster}
        \begin{tabular}{ l |  c | c  c | c }
        \toprule[1pt]
            \multirow{2}{*}{\textbf{Methods}} & \multirow{2}{*}{\textbf{Clustering}} & \multicolumn{2}{c|}{\textbf{CTR}} & \textbf{Time}\\  
            &  & \textbf{AUC} & \textbf{GAUC} & \textbf{Complexity}\\
            \midrule[0.5pt]
            DIN & --- & 0.71739 & 0.67864 & --- \\
        \midrule[0.5pt]
            \multirow{4}{*}{ENCODE}
                & Random & 0.72188 & 0.68484 & $O(K)$\\ 
                & Dy. Routing & 0.72234 & 0.68517 & $O(TKLm)$\\
                & AGG. & \underline{0.72256} & \underline{0.68532} & $O(L^3m)$ \\
                & KMeans  & \textbf{0.72312} & \textbf{0.68626} & $O(TKLm)$\\
        \bottomrule[1pt]
    \end{tabular}
    \end{center}  
\end{table}

\subsubsection{Extend to Different Clustering Algorithm~\textbf{(RQ6)}} To explore whether our proposed ENCODE can be extended to other clustering algorithms, \revise{we experimented the following four algorithms: \textbf{(1) Random}, a random clustering algorithm; \textbf{(2) Dynamic Routing}, as utilized in MIND~\cite{li2019multi}, known for effectively extracting multi-interests; \textbf{(3) Agglomerative}, a bottom-up hierarchical clustering algorithm; \textbf{(4) KMeans}, a straightforward yet effective clustering algorithm. To ensure a fair comparison, we set the number of clusters to $30$. Additionally, to make the clustering algorithm satisfy \textbf{R2}, we uniformly adjusted the distance metric to scaled cosine similarity~(\ie, $sim(\cdot)$ in Eq.~(6)).  The results and time complexity of four clustering algorithms are reported in Table~\ref{tab:ablation_cluster}. 
}

\revise{
Based on the results, it is evident that our ENCODE consistently outperforms DIN by a significant margin when employing different clustering algorithms. 
Even when utilizing a random algorithm for clustering, the outcomes remain favorable, evidenced by a 0.45\% gain in CTR AUC compared to DIN. 
We believe that this is because the attention module for offline multi-interest extraction reduces the sensitivity of the final performance to the clustering results, thereby enhancing the robustness of offline interest representations.
This substantiates the viability of extending ENCODE to incorporate more advanced clustering algorithms.
}




\section{Conclusion}


In this paper, we outline two fundamental prerequisites for long-term sequence modeling and introduce an efficient two-stage model for capturing user interests, called ENCODE, for long-term user behavior modeling. ENCODE achieves state-of-the-art performance maintaining an acceptable level of overhead, striking a balance between performance and efficiency. Moving forward, we are going to (1) design specific training objectives~(\eg, contrastive loss) to further strengthen the interest representation; and (2) design an efficient deep cluster method for large-scale data. 

\section*{Acknowledgments} 

This work was supported by the National Key Research \& Development Project of China (2021ZD0110700), the National Natural Science Foundation of China (62337001) and the Fundamental Research Funds for the Central Universities(226202400058).

\bibliographystyle{unsrt} 
\bibliography{tkde_encode}

\begin{thebibliography}{10}

\bibitem{pi2019practice}
Qi~Pi, Weijie Bian, Guorui Zhou, Xiaoqiang Zhu, and Kun Gai.
\newblock Practice on long sequential user behavior modeling for click-through rate prediction.
\newblock In {\em ACM SIGKDD}, pages 2671--2679, 2019.

\bibitem{pi2020search}
Qi~Pi, Guorui Zhou, Yujing Zhang, Zhe Wang, Lejian Ren, Ying Fan, Xiaoqiang Zhu, and Kun Gai.
\newblock Search-based user interest modeling with lifelong sequential behavior data for click-through rate prediction.
\newblock In {\em ACM CIKM}, pages 2685--2692, 2020.

\bibitem{Qin0WJF020}
Jiarui Qin, Weinan Zhang, Xin Wu, Jiarui Jin, Yuchen Fang, and Yong Yu.
\newblock User behavior retrieval for click-through rate prediction.
\newblock In {\em ACM SIGIR}, pages 2347--2356, 2020.

\bibitem{wu2020sse}
Liwei Wu, Shuqing Li, Cho-Jui Hsieh, and James Sharpnack.
\newblock Sse-pt: Sequential recommendation via personalized transformer.
\newblock In {\em ACM RecSys}, pages 328--337, 2020.

\bibitem{de2021transformers4rec}
Gabriel de~Souza Pereira~Moreira, Sara Rabhi, Jeong~Min Lee, Ronay Ak, and Even Oldridge.
\newblock Transformers4rec: Bridging the gap between nlp and sequential/session-based recommendation.
\newblock In {\em ACM RecSys}, pages 143--153, 2021.

\bibitem{RenQF0ZBZXYZG19}
Kan Ren, Jiarui Qin, Yuchen Fang, Weinan Zhang, Lei Zheng, Weijie Bian, Guorui Zhou, Jian Xu, Yong Yu, Xiaoqiang Zhu, and Kun Gai.
\newblock Lifelong sequential modeling with personalized memorization for user response prediction.
\newblock In {\em ACM SIGIR}, pages 565--574, 2019.

\bibitem{zhou2021large}
Xin Zhou and Yang Li.
\newblock Large-scale modeling of mobile user click behaviors using deep learning.
\newblock In {\em ACM RecSys}, pages 473--483, 2021.

\bibitem{hansen2020contextual}
Casper Hansen, Christian Hansen, Lucas Maystre, Rishabh Mehrotra, Brian Brost, Federico Tomasi, and Mounia Lalmas.
\newblock Contextual and sequential user embeddings for large-scale music recommendation.
\newblock In {\em ACM RecSys}, pages 53--62, 2020.

\bibitem{TaoLLCWCLL22}
Wanjie Tao, Yu~Li, Liangyue Li, Zulong Chen, Hong Wen, Peilin Chen, Tingting Liang, and Quan Lu.
\newblock Sminet: State-aware multi-aspect interests representation network for cold-start users recommendation.
\newblock In {\em AAAI}, pages 8476--8484, 2022.

\bibitem{XuDLHL20}
Xiao Xu, Fang Dong, Yanghua Li, Shaojian He, and Xin Li.
\newblock Contextual-bandit based personalized recommendation with time-varying user interests.
\newblock In {\em AAAI}, pages 6518--6525, 2020.

\bibitem{zhang2021deep}
Weinan Zhang, Jiarui Qin, Wei Guo, Ruiming Tang, and Xiuqiang He.
\newblock Deep learning for click-through rate estimation.
\newblock In {\em IJCAI}, 2021.

\bibitem{ren2019lifelong}
Kan Ren, Jiarui Qin, Yuchen Fang, Weinan Zhang, Lei Zheng, Weijie Bian, Guorui Zhou, Jian Xu, Yong Yu, Xiaoqiang Zhu, et~al.
\newblock Lifelong sequential modeling with personalized memorization for user response prediction.
\newblock In {\em ACM SIGIR}, pages 565--574, 2019.

\bibitem{cao2022sampling}
Yue Cao, Xiaojiang Zhou, Jiaqi Feng, Peihao Huang, Yao Xiao, Dayao Chen, and Sheng Chen.
\newblock Sampling is all you need on modeling long-term user behaviors for {CTR} prediction.
\newblock In {\em ACM CIKM}, pages 2974--2983, 2022.

\bibitem{Rendle10}
Steffen Rendle.
\newblock Factorization machines.
\newblock In {\em IEEE ICDM}, pages 995--1000, 2010.

\bibitem{PanXRZPSL18}
Junwei Pan, Jian Xu, Alfonso~Lobos Ruiz, Wenliang Zhao, Shengjun Pan, Yu~Sun, and Quan Lu.
\newblock Field-weighted factorization machines for click-through rate prediction in display advertising.
\newblock In {\em WWW}, pages 1349--1357, 2018.

\bibitem{Cheng0HSCAACCIA16}
Heng{-}Tze Cheng, Levent Koc, Jeremiah Harmsen, Tal Shaked, Tushar Chandra, Hrishi Aradhye, Glen Anderson, Greg Corrado, Wei Chai, Mustafa Ispir, Rohan Anil, Zakaria Haque, Lichan Hong, Vihan Jain, Xiaobing Liu, and Hemal Shah.
\newblock Wide {\&} deep learning for recommender systems.
\newblock In {\em ACM RecSys}, 2016.

\bibitem{XiaoY0ZWC17}
Jun Xiao, Hao Ye, Xiangnan He, Hanwang Zhang, Fei Wu, and Tat{-}Seng Chua.
\newblock Attentional factorization machines: Learning the weight of feature interactions via attention networks.
\newblock In {\em IJCAI}, pages 3119--3125, 2017.

\bibitem{GuoTYLH17}
Huifeng Guo, Ruiming Tang, Yunming Ye, Zhenguo Li, and Xiuqiang He.
\newblock Deepfm: {A} factorization-machine based neural network for {CTR} prediction.
\newblock In {\em IJCAI}, pages 1725--1731, 2017.

\bibitem{QuCRZYWW16}
Yanru Qu, Han Cai, Kan Ren, Weinan Zhang, Yong Yu, Ying Wen, and Jun Wang.
\newblock Product-based neural networks for user response prediction.
\newblock In {\em IEEE ICDM}, pages 1149--1154, 2016.

\bibitem{LianZZCXS18}
Jianxun Lian, Xiaohuan Zhou, Fuzheng Zhang, Zhongxia Chen, Xing Xie, and Guangzhong Sun.
\newblock xdeepfm: Combining explicit and implicit feature interactions for recommender systems.
\newblock In {\em ACM SIGKDD}, pages 1754--1763, 2018.

\bibitem{wang2017dcn}
Ruoxi Wang, Bin Fu, Gang Fu, and Mingliang Wang.
\newblock Deep {\&} cross network for ad click predictions.
\newblock In {\em ACM SIGKDD}, pages 12:1--12:7, 2017.

\bibitem{CovingtonAS16}
Paul Covington, Jay Adams, and Emre Sargin.
\newblock Deep neural networks for youtube recommendations.
\newblock In {\em ACM RecSys}, pages 191--198, 2016.

\bibitem{HidasiKBT15}
Bal{\'{a}}zs Hidasi, Alexandros Karatzoglou, Linas Baltrunas, and Domonkos Tikk.
\newblock Session-based recommendations with recurrent neural networks.
\newblock In {\em ICLR}, 2016.

\bibitem{ZhouMFPBZZG19}
Guorui Zhou, Na~Mou, Ying Fan, Qi~Pi, Weijie Bian, Chang Zhou, Xiaoqiang Zhu, and Kun Gai.
\newblock Deep interest evolution network for click-through rate prediction.
\newblock In {\em AAAI}, pages 5941--5948, 2019.

\bibitem{FengLSWSZY19}
Yufei Feng, Fuyu Lv, Weichen Shen, Menghan Wang, Fei Sun, Yu~Zhu, and Keping Yang.
\newblock Deep session interest network for click-through rate prediction.
\newblock In {\em IJCAI}, pages 2301--2307, 2019.

\bibitem{SunLWPLOJ19}
Fei Sun, Jun Liu, Jian Wu, Changhua Pei, Xiao Lin, Wenwu Ou, and Peng Jiang.
\newblock Bert4rec: Sequential recommendation with bidirectional encoder representations from transformer.
\newblock In {\em ACM CIKM}, pages 1441--1450, 2019.

\bibitem{zhou2018deep}
Guorui Zhou, Xiaoqiang Zhu, Chenru Song, Ying Fan, Han Zhu, Xiao Ma, Yanghui Yan, Junqi Jin, Han Li, and Kun Gai.
\newblock Deep interest network for click-through rate prediction.
\newblock In {\em ACM SIGIR}, pages 1059--1068, 2018.

\bibitem{li2019multi}
Chao Li, Zhiyuan Liu, Mengmeng Wu, Yuchi Xu, Huan Zhao, Pipei Huang, Guoliang Kang, Qiwei Chen, Wei Li, and Dik~Lun Lee.
\newblock Multi-interest network with dynamic routing for recommendation at tmall.
\newblock In {\em ACM CIKM}, pages 2615--2623, 2019.

\bibitem{jiang2020aspect}
Hao Jiang, Wenjie Wang, Yinwei Wei, Zan Gao, Yinglong Wang, and Liqiang Nie.
\newblock What aspect do you like: Multi-scale time-aware user interest modeling for micro-video recommendation.
\newblock In {\em ACM MM}, pages 3487--3495, 2020.

\bibitem{cen2020controllable}
Yukuo Cen, Jianwei Zhang, Xu~Zou, Chang Zhou, Hongxia Yang, and Jie Tang.
\newblock Controllable multi-interest framework for recommendation.
\newblock In {\em ACM SIGKDD}, pages 2942--2951, 2020.

\bibitem{chen2022efficient}
Qiwei Chen, Yue Xu, Changhua Pei, Shanshan Lv, Tao Zhuang, and Junfeng Ge.
\newblock Efficient long sequential user data modeling for click-through rate prediction.
\newblock {\em arXiv}, abs/2209.12212, 2022.

\bibitem{chang2023twin}
Jianxin Chang, Chenbin Zhang, Zhiyi Fu, Xiaoxue Zang, Lin Guan, Jing Lu, Yiqun Hui, Dewei Leng, Yanan Niu, Yang Song, et~al.
\newblock Twin: Two-stage interest network for lifelong user behavior modeling in ctr prediction at kuaishou.
\newblock In {\em ACM SIGKDD}, pages 3785--3794, 2023.

\bibitem{alexandr2015practical}
Alexandr Andoni, Piotr Indyk, Thijs Laarhoven, Ilya~P. Razenshteyn, and Ludwig Schmidt.
\newblock Practical and optimal {LSH} for angular distance.
\newblock In {\em NeurIPS}, pages 1225--1233, 2015.

\bibitem{charikar2002similarity}
Moses~S Charikar.
\newblock Similarity estimation techniques from rounding algorithms.
\newblock In {\em ACM STOC}, pages 380--388, 2002.

\bibitem{xing2002distance}
Eric Xing, Michael Jordan, Stuart~J Russell, and Andrew Ng.
\newblock Distance metric learning with application to clustering with side-information.
\newblock In {\em NeurIPS}, 2002.

\bibitem{schroff2015facenet}
Florian Schroff, Dmitry Kalenichenko, and James Philbin.
\newblock Facenet: A unified embedding for face recognition and clustering.
\newblock In {\em CVPR}, pages 815--823, 2015.

\bibitem{julian2015image}
Julian~J. McAuley, Christopher Targett, Qinfeng Shi, and Anton van~den Hengel.
\newblock Image-based recommendations on styles and substitutes.
\newblock In {\em ACM SIGIR}, pages 43--52, 2015.

\bibitem{lin2022sparse}
Qianying Lin, Wen{-}Ji Zhou, Yanshi Wang, Qing Da, Qing{-}Guo Chen, and Bing Wang.
\newblock Sparse attentive memory network for click-through rate prediction with long sequences.
\newblock In {\em ACM CIKM}, pages 3312--3321, 2022.

\bibitem{harper2015movielens}
F~Maxwell Harper and Joseph~A Konstan.
\newblock The movielens datasets: History and context.
\newblock {\em ACM Transactions on Interactive Intelligent Systems (TIIS)}, 5(4):1--19, 2015.

\bibitem{kingma2014adam}
Diederik~P Kingma and Jimmy Ba.
\newblock Adam: A method for stochastic optimization.
\newblock In {\em arXiv}, 2014.

\bibitem{sohn2016improved}
Kihyuk Sohn.
\newblock Improved deep metric learning with multi-class n-pair loss objective.
\newblock {\em NeurIPS}, 2016.

\end{thebibliography}

\begin{IEEEbiography}[{\includegraphics[width=1in,height=1.25in,clip,keepaspectratio]{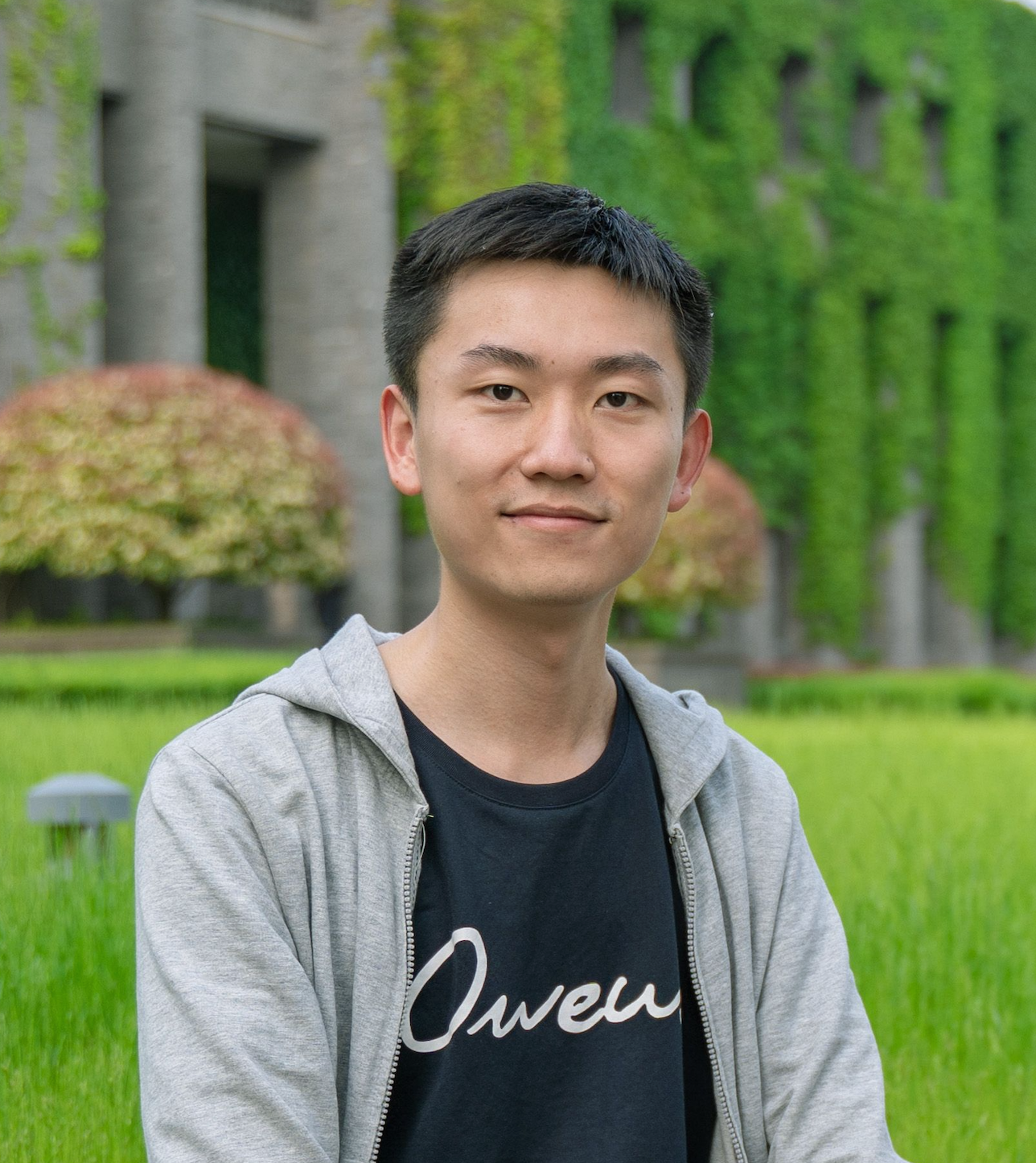}}]{Wen-Ji Zhou} received the BSc and MSc degree in computer science from Nanjing University (NJU), Nanjing, China, in 2016 and 2019, respectively. He is currently a senior algorithm engineer in the search algorithm team of the Department of Alibaba International Digital Commerce Group (AIDC). His research interests are machine learning, recommendation systems, information retrieval and reinforcement learning.
\end{IEEEbiography}

\vspace{-4em}

\begin{IEEEbiography}[{\includegraphics[width=1in,height=1.25in,clip,keepaspectratio]{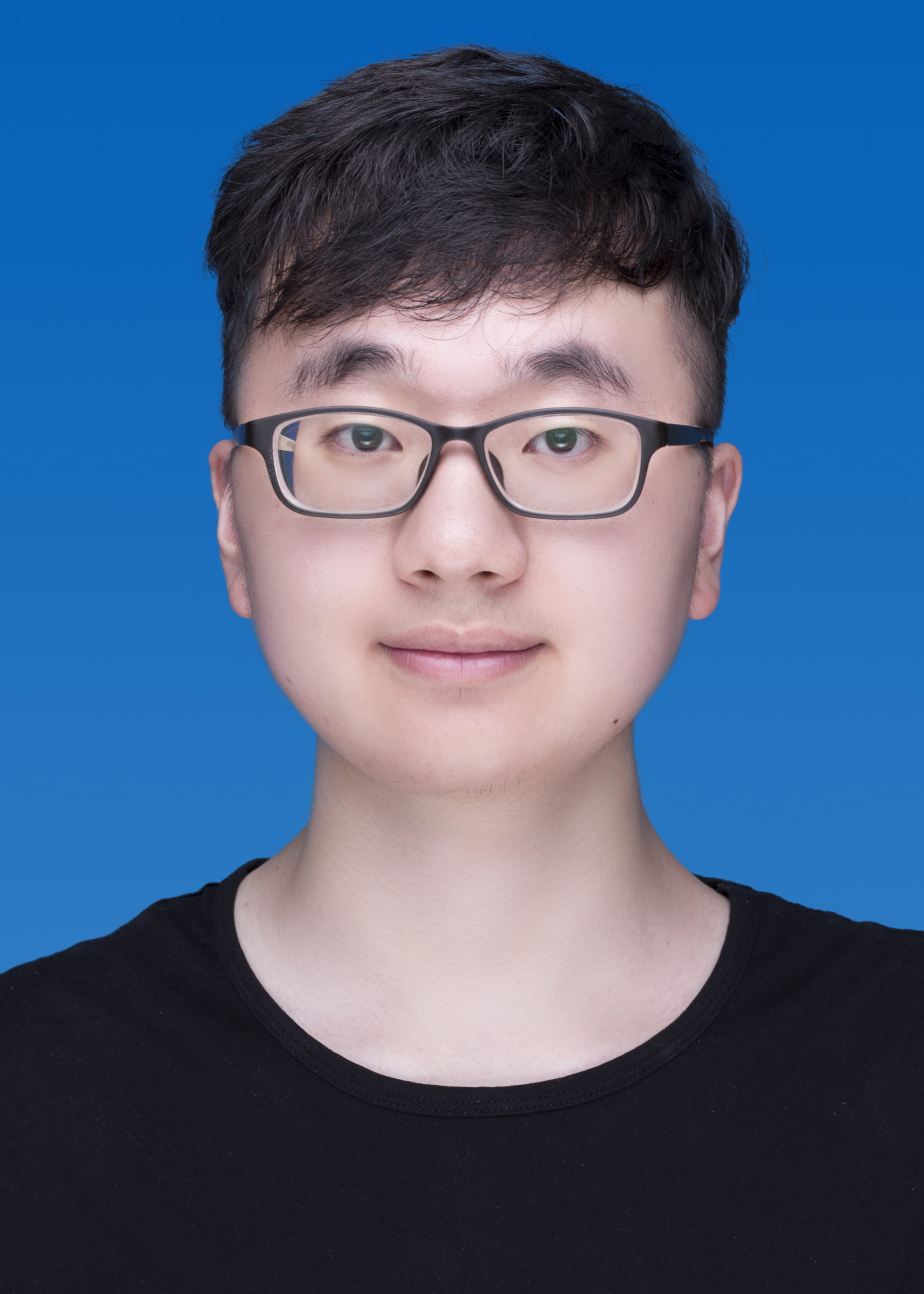}}]{Yuhang Zheng} received the Bachelor (2020)
and Master degree (2023)  in computer science from Zhejiang University, Hangzhou, China. He is currently an algorithm engineer at Alibaba International Digital Commerce Group~(AIDC) in Hangzhou, China. His research interests include recommendation systems, information retrieval, and machine learning.
\end{IEEEbiography}

\vspace{-4em}

\begin{IEEEbiography}[{\includegraphics[width=1in,height=1.25in,clip,keepaspectratio]{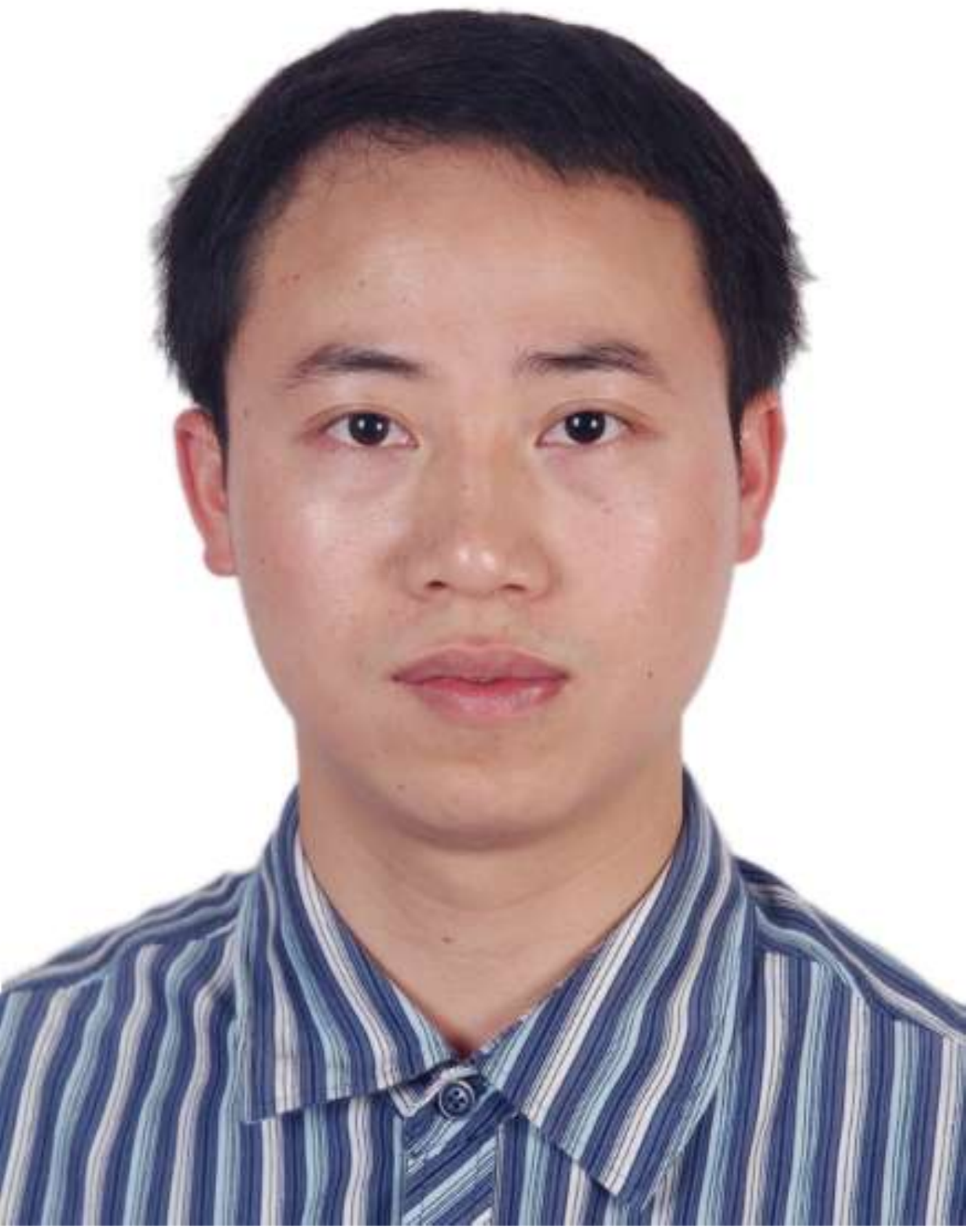}}] {Yinfu Feng} received the Ph.D. degree in computer science and technology from the College of Computer Science, Zhejiang University (ZJU),
Hangzhou, China, in 2015. He is currently a senior staff algorithm engineer in the search algorithm team of the Department of Alibaba International Digital Commerce Group (AIDC). His research interests are computer vision, machine learning and information retrieval.
\end{IEEEbiography}

\vspace{-4em}

\begin{IEEEbiography}[{\includegraphics[width=1in,height=1.25in,clip,keepaspectratio]{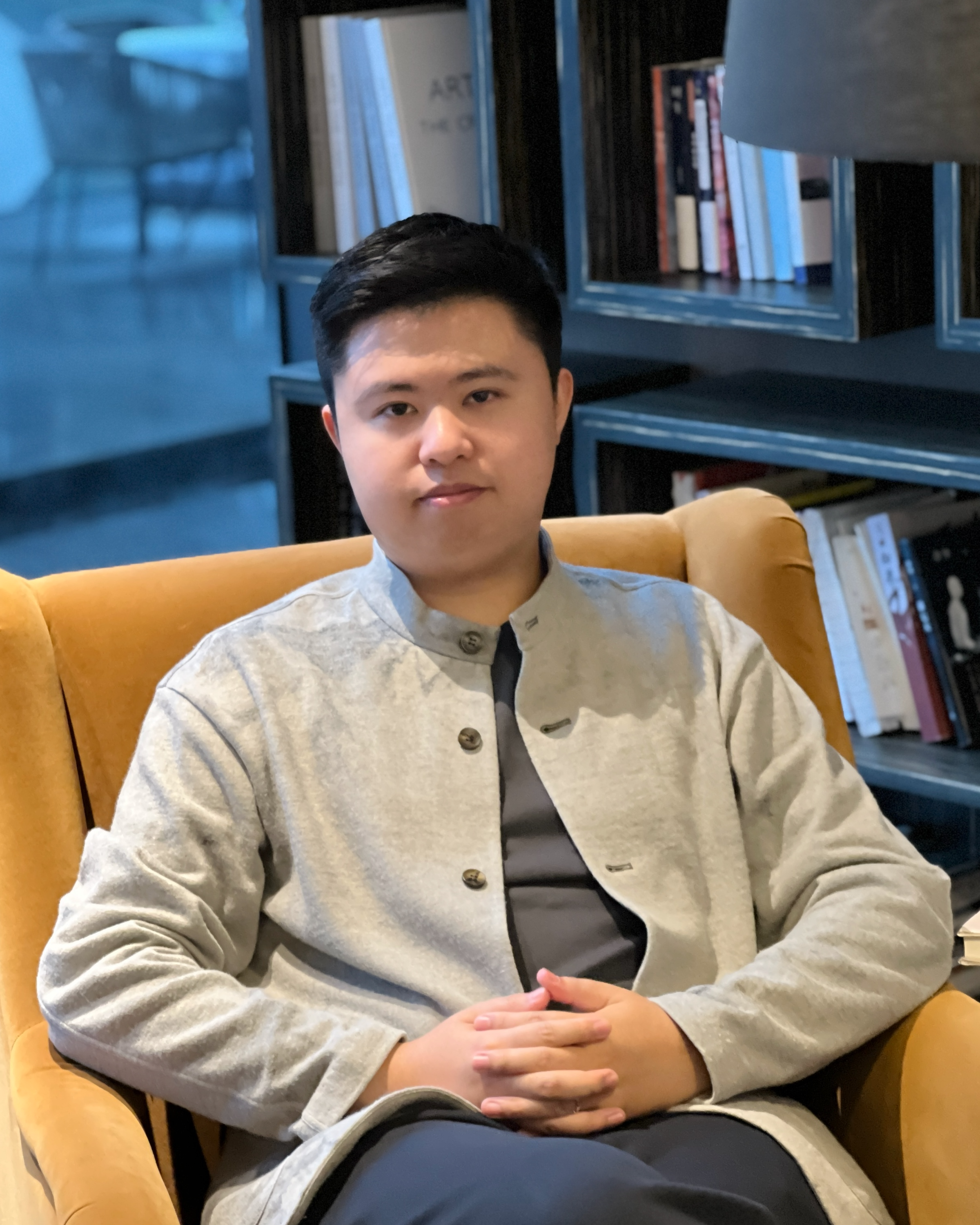}}] {Yunan Ye} received the Ph.D. degree in computer science and technology from the College of Computer Science, Zhejiang University (ZJU), Hangzhou, China, in 2021. He is currently a senior algorithm engineer in the search algorithm team of the Department of Alibaba International Digital Commerce Group (AIDC). His research interests include recommendation systems, machine learning and information retrieval.
\end{IEEEbiography}

\vspace{-4em}

\begin{IEEEbiography}[{\includegraphics[width=1in,height=1.25in,clip,keepaspectratio]{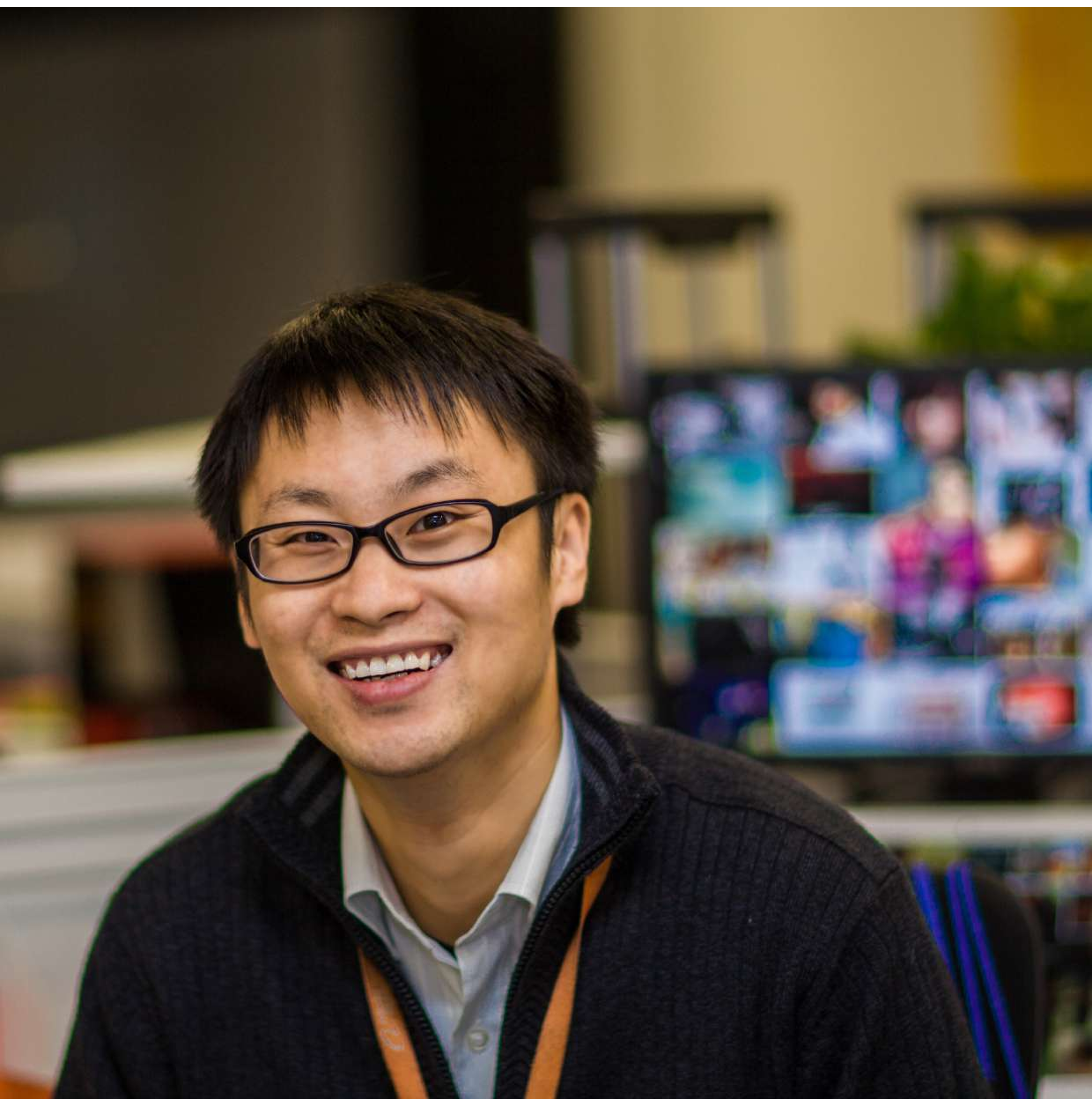}}]{Rong Xiao} received the Master degree in computer science and technology from the College of Computer Science, South Central University for Nationalities, Wuhan, China, in 2011. He is currently a senior staff algorithm engineer in the search algorithm team of the Department of Alibaba International Digital Commerce Group (AIDC). His research interests are machine learning and information retrieval.
\end{IEEEbiography}

\vspace{-4em}

\begin{IEEEbiography}[{\includegraphics[width=1in,height=1.25in,clip,keepaspectratio]{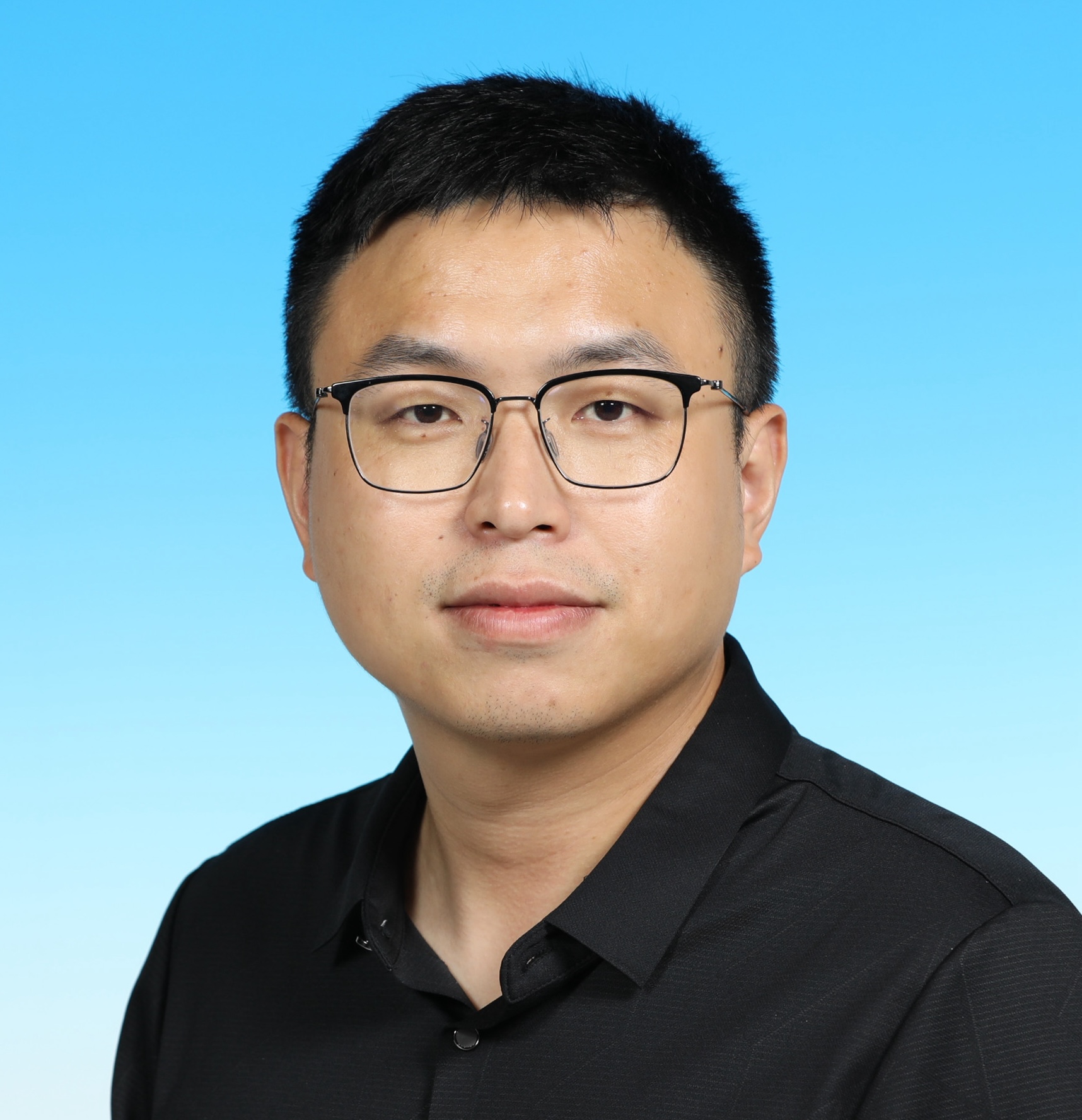}}]{Long Chen} received the Ph.D. degree in Computer Science from Zhejiang University in 2020, and the B.Eng. degree in Electrical Information Engineering from Dalian University of Technology in 2015. He is currently an assistant professor at The Hong Kong University of Science and Technology (HKUST). He was a postdoctoral research scientist at Columbia University and a senior researcher at Tencent AI Lab. His research interests are computer vision, machine learning, and multimedia.
\end{IEEEbiography}

\vspace{-4em}

\begin{IEEEbiography}[{\includegraphics[width=1in,height=1.25in,clip,keepaspectratio]{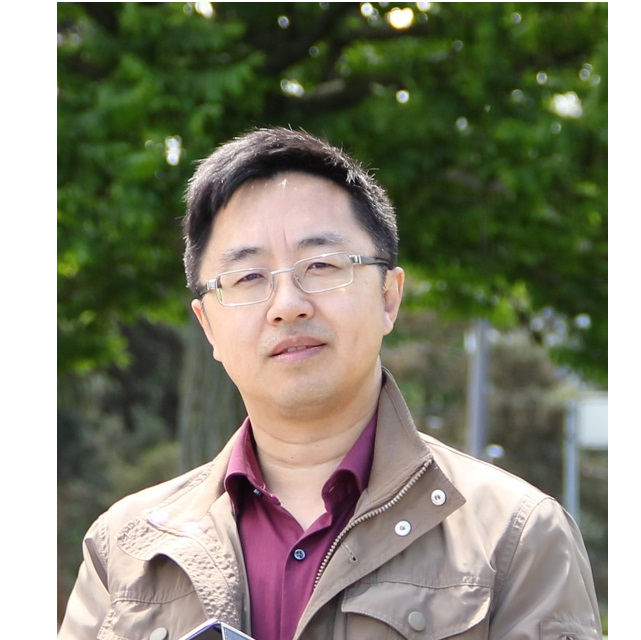}}]{Xiaosong Yang} received the Bachelor (1993) and Master Degree (1996) in Computer Science from Zhejiang University and Ph.D degree (2000) in computing mechanics from Dalian University of Technology, China.  He is currently Professor of Computer Animation, Deputy Head of Department in the National Centre for Computer Animation, Bournemouth Unviersity, United Kingdom. His current research interests include computer graphics, computer vision, machine learning, data mining, digital health and virtual reality. 
 
\end{IEEEbiography}

\vspace{-4em}

\begin{IEEEbiography}[{\includegraphics[width=1in,height=1.25in,clip,keepaspectratio]{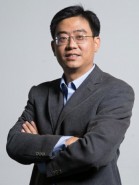}}]{Jun Xiao} received the Ph.D. degree in computer science and technology from the College of Computer Science, Zhejiang University, Hangzhou, China, in 2007. He is currently a professor with the College of Computer Science, Zhejiang University. His current research interests include cross-media analysis, computer vision and machine learning.
 
\end{IEEEbiography}

\end{document}